# AI as a Teaching Partner: Early Lessons from Classroom Codesign with Secondary Teachers

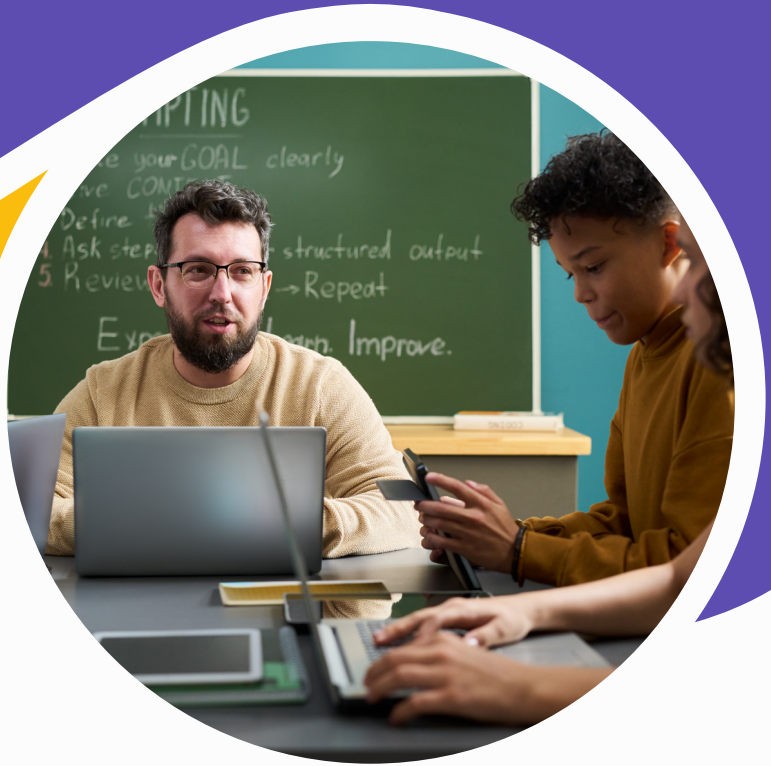

**Authors:**

Alex Liu[1], Lief Esbenshade[1], Shawon Sarkar[1], Zewei (Victor) Tian[1], Min Sun[1], Zachary Zhang[2], Thomas Han[2], Yulia Lapicus[2], Kevin He[2]

[1] University of Washington, [2] Colleague AI

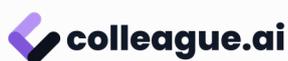 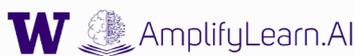 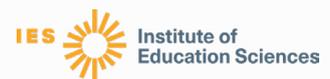

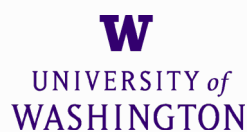 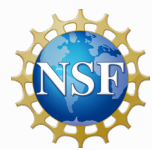

Acknowledgments: This work is supported by the Institute of Education Sciences of the U.S. Department of Education, through Grant R305C240012 and by several awards from the National Science Foundation (NSF #2043613, 2300291, 2405110) to the University of Washington, and a NSF SBIR/STTR award to Hensun Innovation LLC (#2423365). The opinions expressed are those of the authors and do not represent views of the funders.

# Table of Contents

# Executive Summary

This report presents a comprehensive account of the Colleague AI Classroom pilot, a collaborative design (co-design) study that brought generative AI technology directly into real classrooms. In this study, AI functioned as a third agent, an active participant that mediated feedback, supported inquiry, and extended teachers' instructional reach while preserving human judgment and teacher authority.

Over seven weeks in spring 2025, 21 in-service teachers from four Washington State public school districts and one independent school integrated four AI-powered features of the Colleague AI Classroom into their instruction: **Teaching Aide, Assessment & AI Grading, AI Tutor, and Student Growth Insights.** More than 600 students in grades 6-12 used the platform in class at the direction of their teachers, who designed and facilitated the AI activities.

During the Classroom pilot, teachers were co-design partners: they planned activities, implemented them with students, and provided weekly reflections on AI's role in classroom settings. The teachers' feedback guided iterative improvements for Colleague AI. The research team captured rich data through surveys, planning and reflection forms, group meetings, one-on-one interviews, and platform usage logs to understand where AI adds instructional value and where it requires refinement.

## Key Findings

- **Pedagogical judgment and AI competency jointly determine instructional impact.** Effective integration requires more than technical familiarity. Teachers needed both pedagogical judgement and tool fluency to structure student-AI interactions with clear goals, frame prompts, monitor progress, and model responsible use. Those with prior instructional experience using AI, and who perceived AI as instructionally useful prior to the pilot, were more likely to more actively experiment with students, iteratively refine AI-generated feedback, and report confidence in reuse.

- **AI functions as a "third agent" when teachers provide scaffolds and framing.** AI-supported features were most effective when teachers provided clear framing and instructions for both students and AI. Goal-oriented prompts (e.g., seminar preparation, proof sketches) led to deeper student-AI exchanges. In practice, AI acted as a "third agent" in the classroom, an additional instructional presence alongside teacher and student. AI proved especially useful for learning in guided exploratory preview and formative feedback cycles, assisting students with creating an initial draft, providing feedback, and prompting revision, while teachers retained final evaluative authority.

- **Implementation patterns vary by grade band and subject.** Teachers' use of AI reflected both the age group they served and their subject area. Middle-grade teachers (6-8) relied on clear workflows, shorter outputs, and strong monitoring, while high school teachers (9-12) made greater use of sustained dialogue with AI. In math and science, trust hinged on accurate, domain-specific feedback, whereas ELA and social studies teachers reported that off-topic or generic responses undermined learning value.

# Main Features in Colleague AI Classroom

**Teaching Aide:** The AI Teaching Aide functions as a classroom discussion partner for students, designed and guided by teachers. It supports student inquiry through structured, teacher-authored prompts that scaffold learning, reinforce key concepts, and sustain engagement during class activities. Students engage in one-on-one conversations with the AI; all students see the same initial message from the AI system and conversations automatically tailor themselves to the student's responses, while staying aligned to the teacher's instructions. Teachers can preview the tool as students, join live conversations, and review summaries that highlight common themes, misconceptions, and emerging needs. By mediating dialogue rather than directing it, the Teaching Aide extends teachers' instructional reach while preserving their professional oversight and judgment.

**Assessment with AI Feedback:** Colleague AI's assessment tools combine automated rubric generation with AI-assisted grading to make classroom evaluation faster, fairer, and more formative. Teachers can align assessments with national or state standards (e.g., CCSS, NGSS), select instructional frameworks such as Bloom's Taxonomy or Webb's Depth of Knowledge, and edit every criterion to ensure accuracy and contextual fit. The AI applies these rubrics to student work, producing instant narrative feedback that highlights strengths, identifies misconceptions, and suggests next steps, while teachers retain full authority over final grades.

**AI Tutor:** The AI Tutor provides students with on-demand academic support for homework, test preparation, and project-based learning under teacher supervision. Designed for guided independence, it helps students clarify concepts, practice reasoning, and receive formative feedback without giving away answers. All conversations are saved and summarized for teacher review, allowing educators to monitor progress, identify misconceptions, and ensure responsible use. By promoting self-advocacy and critical thinking while maintaining oversight, the AI Tutor extends learning beyond class time without diminishing the teacher's role.

**Student Growth Insights:** Student Growth Insights (SGI) transforms classroom data into actionable feedback for teachers. It aggregates patterns from student–AI interactions, highlighting common questions, misconceptions, and areas of progress at both the class and individual level. Teachers use these insights to differentiate instruction, plan reteaching, and communicate learning growth to parents and colleagues.

A demonstration of the Colleague AI platform's Classroom features is available on [YouTube.](#) This video incorporates learning from the codesign sessions to improve the tool.

**Colleague AI Classroom features demonstrated notable potential to reduce teacher burden while enhancing instructional feedback loops.** Features like SGI automatically surfaced class-wide misconceptions and student growth patterns; AI-assisted grading streamlined feedback using teacher-authored rubrics; and the Teaching Aide enabled real-time personalized student support without requiring constant teacher intervention. Teachers reported that Colleague AI Classroom features augmented instructional practices through human-oversighted automation, differentiation, and targeted feedback.

**The pilot suggests early-stage promise for both teacher- and student-level AI-amplified outcomes.** At the teacher level, participants reported faster turnaround times for personalized feedback and growing confidence in leveraging AI for both in-class discussions and formative assessments. At the student level, observed outcomes included better scaffolded inquiry, visible self-improvement behavior in response to feedback, and increased use of AI as a learning partner, particularly when teachers structured prompts or rubrics with clear goals and evaluative expectations. These outcomes, while exploratory, suggest measurable potential for AI-supported instructional improvement with sustained support and infrastructure.

## Toward Effective Integration: Recommendations and Next Steps

**AI adds value when it is structured, scaffolded, and goal-oriented.** Clear prompts, short default replies, and visible end-products (exit tickets, notebook entries, seminar preparation) enabled students to engage thoughtfully with the AI. Without this framing, conversations devolved into aimless back-and-forth or became too overwhelming for younger learners.

**Teacher AI competency drives adoption and impact.** Teachers who entered the pilot with prior experience using AI for instruction, or who strongly believed in its potential, implemented more activities, framed the AI's role more effectively, and expressed higher confidence in reuse. Those without such experience typically conducted only one trial and reported challenges.

**Formative feedback over summative scores.** Across all assessments, teachers and students valued narrative feedback, specific comments that surfaced misconceptions and guided revision, far more than numeric scores, which were seen as inconsistent or opaque. Teachers used the AI as **a first-pass feedback engine** and retained final grading authority.

**Developmental and subject differences matter**. Middle school students benefited from concise outputs, sentence stems, and step-by-step guidance, whereas high school students tolerated deeper, more exploratory exchanges. Humanities and interdisciplinary STEM courses achieved the highest engagement, while math and earth science saw more mixed results due to the AI's difficulty with diagrams and technical reasoning.

**SGI delivers high leverage with little effort.** Student Growth Insights, which aggregate and display students' most common questions, saved teachers hours of scanning and enabled targeted reteaching. Teachers rated SGI as one of the most helpful elements for understanding student needs.

**AI is a partner, not a replacement.** Teachers consistently viewed AI as a third agent that augments teaching and learning rather than supplants it. Students appreciated quick, personalised feedback but wanted teachers to see their work and remain involved. The most successful implementations followed a collaborative model, with teachers framing the activity, AI providing initial support, and teachers interpreting and extending the AI's feedback.

Taken together, this pilot study demonstrates that generative AI can meaningfully support classroom instruction when it is thoughtfully designed, contextualised by teachers, and integrated into existing routines. This report details the study design and results for each feature, offers subject-specific insights, and concludes with practical implications for educators and school leaders.

# Study Design

## Participants and Recruitment

The pilot took place in five Washington State districts—Bellevue, Issaquah, Northshore, Bellingham—and the independent Eastside Preparatory School. 21 in-service teachers participated, spanning middle and high school, with classes ranging from 15 to 33 students. Teachers represented a wide range of disciplines:

- Mathematics (9 classes): Middle school math, Algebra 1, and Geometry
- Science (7 classes): General science, biology, earth science, marine biology, biotechnology, and forensic science
- ELA and Social Studies (10 classes): Literature, English language arts, history, and civics.
- Other (14 classes): business law, design, economics, electrical engineering, marketing, programming, and Spanish.

Teachers were recruited through district partnerships and compensated for their time. Each of the partner districts have signed a Memorandum of Understanding (MOU) with Colleague AI explicitly stating the terms of the research work and including a detailed [privacy policy](). Participation in the study was entirely voluntary. Teachers were not required to use all available features and could opt out of specific activities if they felt they were not appropriate for their students or did not align with their planned curriculum. Similarly, student participation was voluntary. Teachers did not mandate student use of the platform and designed their activities in two parallel versions: one incorporating AI and one without. Students were free to opt out of using AI for any activity or assessment without it affecting their learning or classroom standing. In total, 25 teachers were recruited for the study; four withdrew at the beginning of the study citing time constraints and are not included in the analysis. The study protocol was approved by the University of Washington Institutional Review Board (IRB).

### Research Questions

- (1) How do teachers design and implement AI-assisted tools in typical classroom instruction?
- (2) How does teacher judgment and AI competency shape student-AI learning interactions and instructional outcomes?
- (3) What professional support, feedback, and design conditions enable effective AI integration into teaching?

# Timeline and Teacher Learning Arc

The pilot spanned seven weeks, designed as a professional learning arc that gradually introduced each feature and provided time for planning, implementation, and reflection. Teachers were asked to spend approximately 5-10 hours per week across the study, including attending weekly online meetings, planning AI-integrated lessons, and implementing those lessons in their classrooms. Teachers implemented the Colleague AI tool in their live, synchronous classrooms and during their regular classroom prep time. Teacher participation with the researchers included live sessions for discussion and troubleshooting and asynchronous surveys on platform features.

**Table 1. 7-Week Colleague AI Implementation Timeline**

| Week | Focus Area | Key Activities |
| --- | --- | --- |
| Week 1 | Onboarding | Platform introduction, logistical setup (rosters, consent), group communication channel setup, core feature exploration in demo environment |
| Week 2 | Introducing Teaching Aide | Introducing Teaching Aide feature, designing two student-facing conversations with structured worksheets, curriculum alignment without instructional disruption |
| Week 3 | Preparing Assessment and AI Grading | Assessment feature introduction, rubric generation exploration, AI feedback and grading options, Teaching Aide prompt refinement, assessment planning |
| Week 4 | Implementing Teaching Aide | First student-AI Teaching Aide session in class, conversation monitoring, exit reflection submissions, group meetings on implementation challenges and student responses |
| Week 5 | AI Feedback and Assessment | Two assessment implementations with optional immediate AI feedback, reflective group meeting between implementations, exploring formative vs. summative AI assessment roles |
| Week 6 | Student Growth Insights | Review of AI-generated student activity summaries, examination of common questions and misconceptions, SGI data accuracy evaluation, instructional adjustments based on insights |
| Week 7 | Wrap-Up and Reflection | Implementation lessons synthesis, responsible AI use discussions, post-study survey completion, key takeaways identification for AI-supported teaching and learning |

The progression from technical onboarding to pedagogical reflection emphasizes the success criteria for AI implementation requires both operational readiness and thoughtful consideration of educational impact. Each week builds upon previous learning while introducing new capabilities that support different aspects of teaching and assessment.

# Limitations and Future Research

Findings in this report are based on teacher plans, weekly meetings, and exit tickets. Future research directions include connecting teacher and student usage patterns to actual student artifacts, such as revisions, seminar contributions, and lab reasoning, to better understand learning outcomes and instructional impact. These connections may also help illuminate how students engage with AI suggestions over time, and which types of feedback are most likely to support conceptual understanding.

Several limitations should be noted: the sample was geographically limited to the Puget Sound region, not all participants completed surveys, and the study relied on teacher self-report and platform logs rather than direct classroom observation. In addition, findings are based on a single generative AI platform and may not generalize across tools.

Despite these constraints, the study offers actionable insights. Teachers consistently framed **Teaching Aide's as a teaching assistant or learning partner, not a teacher replacement.** This framing prompted us to rename the tool from AI Discussion to Teaching Aide. They viewed AI-generated narrative feedback as **a formative scaffold, not a summative judge.** Students valued quick, personalized feedback but expected teacher involvement for trust and relational connection. For educational practitioners, this points toward **a teacher-student-AI collaborative model**: AI can extend instructional reach, but educators must remain the final interpreters of student learning.

Several opportunities for future development also emerged. One key area is designing age-appropriate levels of cognitive friction across grade bands. Field feedback suggests that younger learners may benefit from low-friction, scaffolded AI interactions, while older students may require increased cognitive challenge to avoid overreliance and foster deeper thinking. Systematic scaffolding of effort and ambiguity, especially across elementary, middle, and high school use cases, warrants further study.

Additional research could explore the thresholds of teachers' cognitive friction in AI-enriched environments, particularly when processing multimodal inputs from both students and AI tools in real time, and when monitoring student-AI interactions at scale. Simplifying teacher-facing interfaces, embedding just-in-time AI fluency supports, and improving alignment between AI-generated and teacher-assigned feedback are promising strategies to reduce this burden.

These directions point to critical gaps in the current AI in education research landscape: measuring the instructional return on teacher time in AI-supported classrooms, identifying scalable pathways for building teacher AI competency, and understanding how human-AI collaboration shapes learning across developmental stages. Addressing these gaps will be essential for designing responsible, effective, and pedagogically sound AI integration in K-12 education.

The observations from this study underscore the importance of investing first in **teacher AI competency and tool fluency** to maximize the return on investment of school or district AI adoption. With targeted product refinements, AI can reliably augment instruction as a "third agent"--assisting teachers, tutoring students in conversations, accelerating formative feedback cycles, and closing the instructional feedback loop through SGI insights and teacher-facing features.

# Teacher Experience and AI Perspectives

## Experienced Teachers with Limited AI Exposure

At the beginning of the study, participating teachers completed a survey documenting their instructional experience, technology use, and prior exposure to AI. This baseline context frames how teachers approached AI integration during the pilot and highlights the supports they found most necessary.

> **To what extent do you feel knowledgeable and experienced in the following areas?**

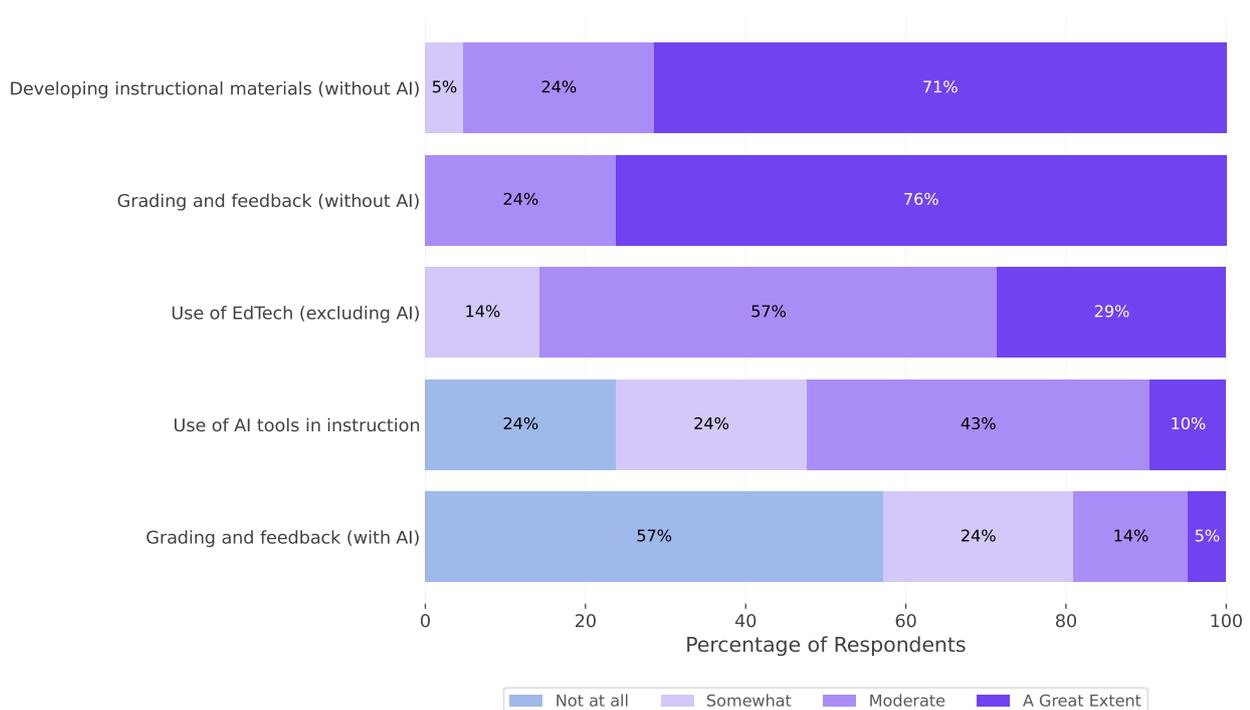

Question N = 21 (pilot participants)

Teachers entered the study with confidence in their foundational instructional skills. Nearly all reported extensive to moderate experience designing lessons (95%) and providing feedback without AI (100%), confirming that the pilot aimed to **layer emerging technologies onto teachers' professional expertise.** In discussion with teachers, it became clear that they were not looking for AI to teach them how to do their jobs; rather, they sought ways for it to enhance, streamline, or expand the instructional work they already performed with skill and consistency.

In contrast, prior experience with AI was limited and uneven. Most teachers reported moderate (57%) to high

(29%) use of non-AI digital tools, such as learning management systems (LMS) and instructional apps, but relatively little direct engagement with AI for teaching (52%). Roughly one-quarter of participants **had never used AI in an instructional capacity.** Those who had experimented with AI tools reported using general-purpose chatbots (e.g., ChatGPT, Claude, Gemini) or creative generators (e.g., MidJourney, Gamma), while only a small subset had explored education-specific platforms (e.g., MagicSchool AI, Class Companion, PackBack). Even then, most described their experience as only "somewhat" or "moderate."

Exposure to AI for grading or feedback was especially low: the majority (57%) reported **no prior use in assessment contexts.** For most participants, this pilot represented their first structured opportunity to implement AI with students in a classroom setting. Curiosity was high, but practical experience, particularly with high-stakes tasks like assessment, remained minimal. A few teachers had previously experimented informally with Colleague AI, but these trials were exploratory and not part of any systematic instructional implementation.

# Teachers Entered the Study Seeing AI as a Backstage Tool, Not a Classroom Partner

**Teacher perceived effectiveness of AI-powered tools in supporting instructional practice across key areas**

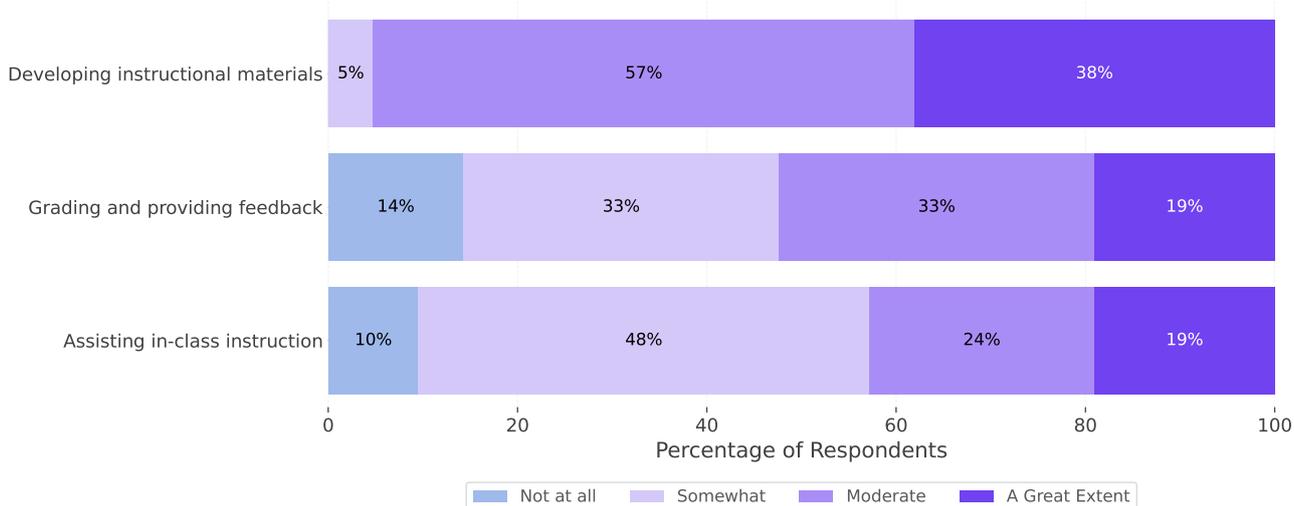

Question N = 21 (pilot participants)

When asked about the potential usefulness of AI, teachers expressed cautious optimism. Most (95%) saw clear value in lesson preparation tasks, such as generating drafts of instructional materials. Perceptions of AI's role in grading and feedback were more mixed, ranging from strong interest in its potential (19%) to skepticism about accuracy and reliability (14%).

The most cautious outlook emerged around AI's presence at in-class instruction. While teachers were open to experimenting with AI as a live classroom support, most remained unconvinced or uncertain (57% selected "Somewhat" or "Not at all" ). Overall, teachers entered the study with **diverse perspectives on AI's promise,** but a common tendency to view it primarily as **a behind-the-scenes assistant** rather than an active agent in the classroom .

# Prior AI Experience Shaped Implementation Across Tools

Teachers' prior experience and beliefs about AI shaped implementation across both Teaching Aide and Assessment features, though the effects varied based on task demands. Assessment, being tied to grading, required stricter fidelity, while Teaching Aide allowed for more flexibility.

Teachers with prior AI experience tended to have stronger first runs, clearer framing, and greater confidence to reuse both tools. In Teaching Aide, they modeled sample chats, tied conversations to class products, and defined clear finish lines. In Assessment, they created focused rubrics, enabled student-visible feedback, and used revise-and-resubmit cycles.

Beliefs about AI's usefulness also predicted depth of use. Teachers who entered the study with stronger belief in AI's ability to support learning or feedback made more intentional instructional moves, whether anchoring Teaching Aide's in seminar prep or using AI feedback to guide student revisions in Assessment compared to teachers who were more skeptical. General edtech comfort reduced logistical friction as well. Teachers more familiar with digital tools encountered fewer problems with navigation, logins, and submission flow, allowing smoother integration of AI into classroom routines.

The nature of the task plays a central role in shaping how AI features are received. Teaching Aide functions more like tutoring, tolerating ambiguity and open-ended exploration, while AI Assessments are anchored in judgment, requiring consistent scoring and transparent alignment to rubrics. Because grading carries higher stakes, any drift in scoring scale or off-rubric deductions can quickly erode trust, even among teachers open to AI use. Tool maturity also differs: Teaching Aide pain points focused on response length, tone, and monitoring, while Assessment challenges centered on rubric enforcement, point scaling, file submission issues, and revision workflows. Subject-specific complexity further compounds the gap. Math and computer science, for example, present greater hurdles for automation than conversational reasoning in ELA or social studies. Finally, cognitive load varies: long AI-generated feedback in Assessment was more overwhelming to students than lengthy AI responses replies, reinforcing the need for "short → expand" controls in both contexts.

## Table 3. Pre-Survey Factors Shaping Teaching Aide and Assessment Use

| Pre-survey factor | Teaching Aide pattern | Assessment pattern | Reason for Pattern Differences |
|---|---|---|---|
| **Prior AI use in instruction** | Boosts engagement via better prompts and norms; Better framing to help students "tolerate" study mode more. | Helps set up revise-and-resubmit, but **does not guarantee trust in AI scores.** | **Grading demands more fidelity** and trust between students and graders. |
| **Belief in AI for grading** | Only weakly related (perceived effectiveness of AI in grading beliefs don't matter much to a dialogue task). | Strong predictor of **actually using AI feedback** and keeping it **student-visible** (teacher is the final grader regardlessly). | Feature-belief match: this belief maps directly to Assessment behaviors. |
| **Subject area** | **ELA/SS/Science:** good fit for idea-building; **Math:** frustration with Socratic style but still usable for concept talk. **CTE:** higher inherent motivation to view AI as technology. | **ELA/Spanish:** most positive (clear writing feedback); **Math/CS: lowest trust** (scale drift, off-rubric deductions, inconsistency), **CTE/Science:** mixed (file/OCR/code issues). | Evaluation in STEM needs **precise rubrics, per-question structure,** and **domain rules;** while dialogue can stay conceptual. |
| **Grade band (MS vs HS)** | MS needs **short replies, stems, visible progress;** HS tolerates productive struggle. | MS is sensitive to **UI clarity** (submit/see feedback/resubmit) and **point-scale consistency;** HS more tolerant if feedback and revision can generate meaningful improvement for their work. | Younger learners are more disrupted by wording length and workflow friction; score inconsistency undermines buy-in faster. |
| **Trust anchor** | **Pedagogical fit** (tone, length, next step). | **Scoring fidelity** (fixed scale, alignment to rubric rows, consistency across reruns). | Different success criteria: conversation quality vs. auditability of evaluation. |

# AI-Assisted Conversation via Teaching Aide

The Teaching Aide[2] feature allows teachers to create structured student-AI conversations by adding an additional layer of prompt instruction for AI. Teachers create custom instructions for the AI on the topic it should discuss and how it should discuss it with students. All students then receive the same initial message from the AI and the conversation automatically adapts to their responses. The goal is to broaden classroom discourse—personalizing learning, prompting deeper thinking, and scaffolding inquiry—without replacing the teacher's voice[3].

In the pilot, 18 teachers created 114 distinct AI-assisted in-class conversations across 21 classrooms in which up to 600 students participated, covering subjects from math to humanities. Three teachers decided not to implement the Teaching Aide tool because it did not fit with their curriculum plan.

**Total number of Teaching Aides Configured by Teachers by Subject and Grade Level**

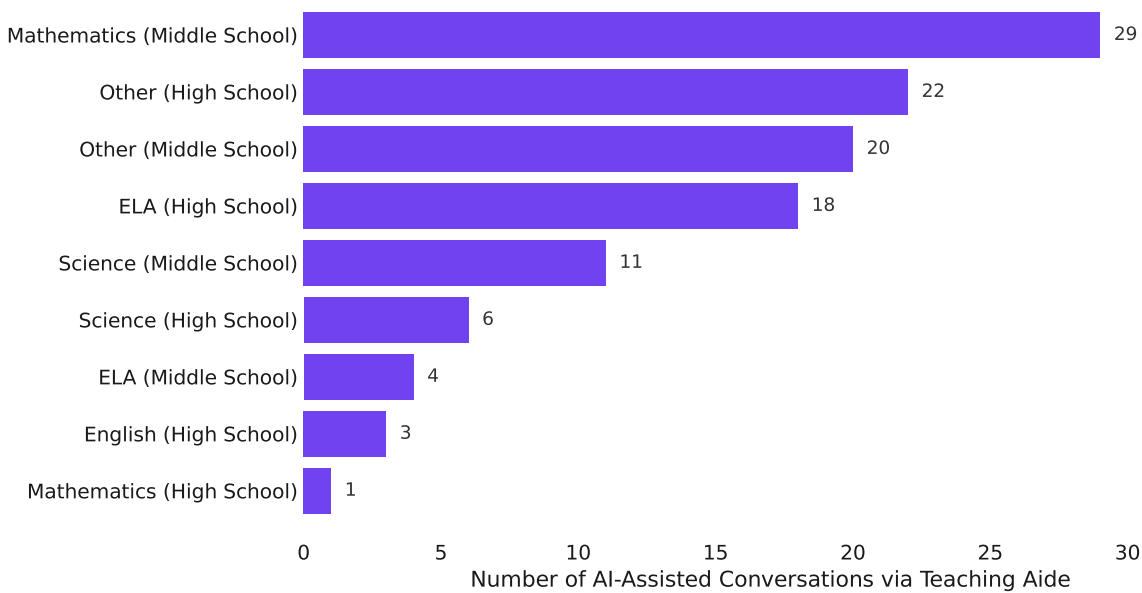

Note: This chart reflects the number of Teaching Aides that were configured by the 18 teachers who participated in this part of the pilot. Teachers created between 1 and 26 Teaching Aides, with an average of 6.3 per teacher.

---

[2] The tool was originally referred to as 'AI Discussion'. Based on feedback during the codesign sessions, it was renamed to Teaching Aide.
[3] A demonstration of the Teaching Aide tool is available on YouTube. This video incorporates learning from the codesign sessions to improve the tool.

# Teacher Framing and Tool Design Drive Meaningful Student-AI Conversations

**Teacher rated student engagement with Teaching Aide Conversations**

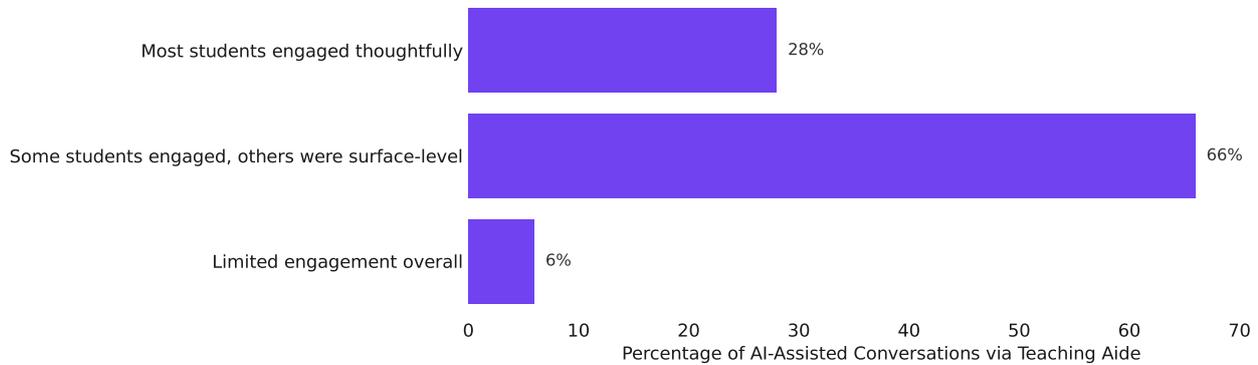

Note: Responses collected from 18 teachers. Weights normalized to 1 for teachers who reported multiple classes.

Approximately 28% of teachers reported classroom implementations with high levels of engagement, with most students participating thoughtfully in their AI-assisted conversations. Around 66% of teachers experienced mixed engagement in their classrooms, while only 6% reported low overall engagement. Teachers played a critical role in guiding successful implementation. They modeled effective prompts, actively circulated during sessions, and consistently framed the AI as a thought partner rather than a search engine. In classrooms where teachers were less present or involved, students were more likely to treat the AI as a source of entertainment or disengage entirely. Teachers also found that tying Teaching Aide sessions to graded assignments (e.g. as a tool to support the student's use of visual imagery in a Spanish writing assignment) helped promote thoughtful participation, especially among younger students or during the early stages of implementation.

## Engagement Patterns Varied by Grade Level and Subject

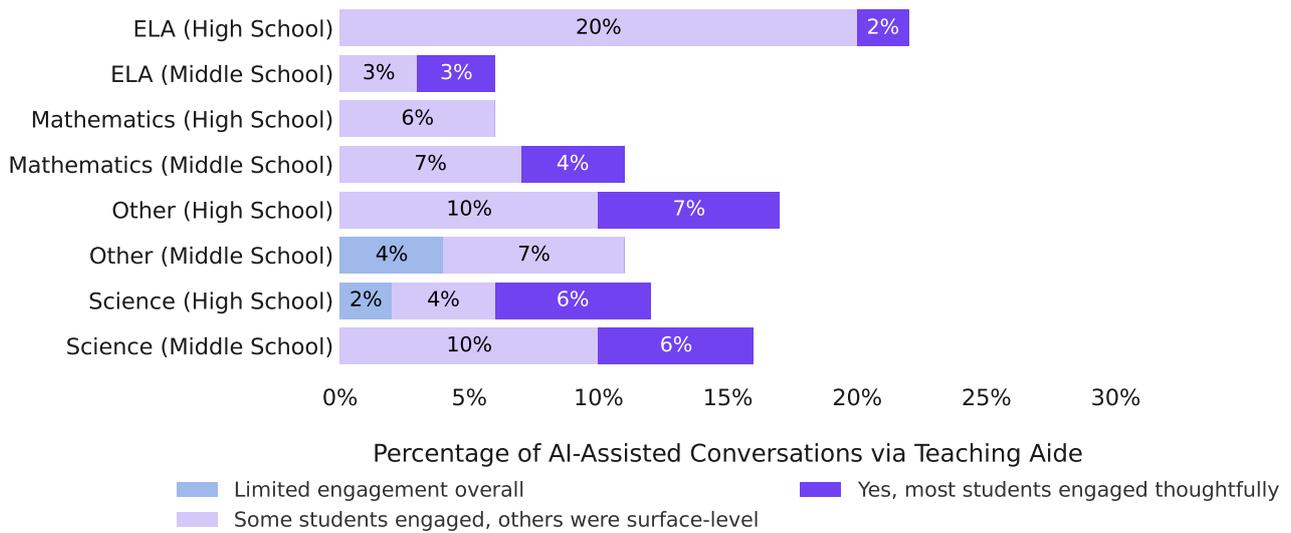

Note: Responses collected from 18 teachers. Weights normalized to 1 for teachers who reported multiple classes.

Depth of engagement varied noticeably. Middle school math showed more thoughtful, whole-class engagement than high school math (4% vs. 0%), while science was similar across grades (6% each). "Other" subjects—including Spanish, engineering, and interdisciplinary STEM—were stronger at the high school level (7% vs. 0% in middle school). ELA classrooms revealed the sharpest contrasts: high school ELA frequently showed uneven participation (20% surface-level), with only 2% reporting thoughtful engagement compared to 3% in middle school. Overall, younger students benefited most from teacher-initiated conversation starters and explicit participation norms, whereas high school students required additional support to build trust in the AI and move beyond shallow exchanges toward deeper inquiry.

## Student Engagement Changed by the Number of Implementations

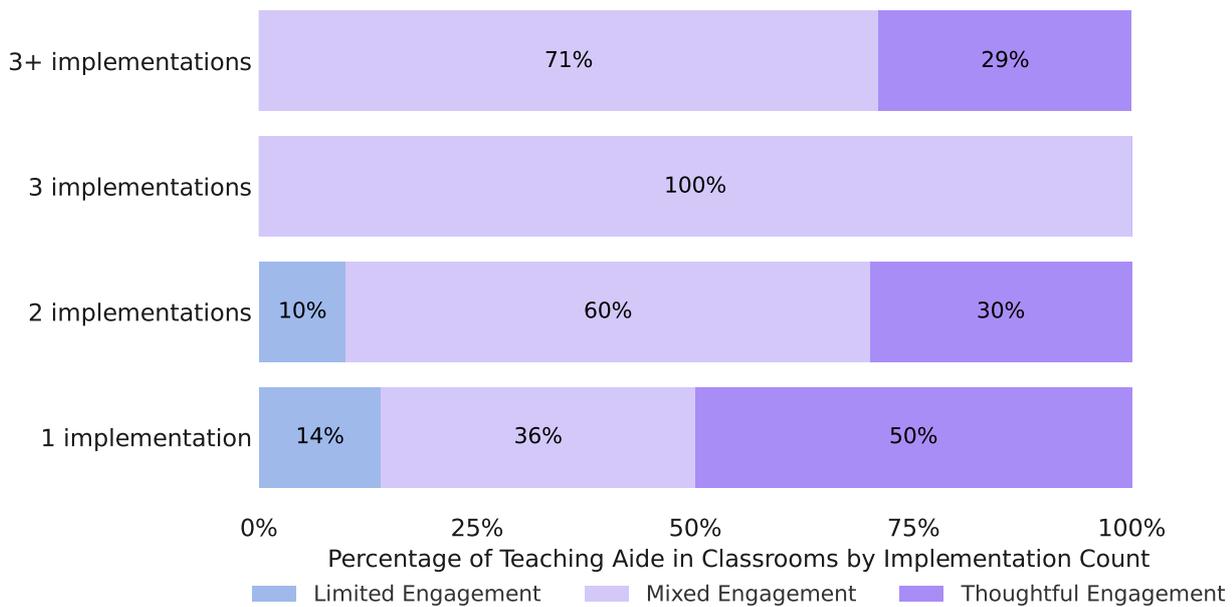

Note: Responses collected from 18 teachers. Weights normalized to 1 for teachers who reported multiple classes.

Classrooms that used the Teaching Aide tool multiple times generally saw higher levels of student engagement. Teachers described learning how to refine prompts, clarify AI conversation norms with students, and troubleshoot workflow friction as they gained experience. In group debriefings teachers who actively participated in brainstorming use cases and strategies for scaffolding participation generally reported stronger results in later rounds of implementation. By contrast, classrooms that ran only a single use of Teaching Aide more often noted challenges such as overly long responses, mismatched complexity, or low student buy-in, which sometimes discouraged further use. While we cannot fully separate whether repeated use caused higher engagement or reflected teachers' willingness to persist, the overall pattern suggests that iteration and collaborative planning supported more meaningful outcomes.

While teacher framing was essential, the design of the AI tool itself also contributed to successful engagement. Key design elements included explicit finish lines to help bring conversations to a meaningful close, a clear display of the teacher's initial instructions at the top of the session, and real-time monitoring dashboards that allowed teachers to stay connected to student thinking during AI conversations. Without these supports, conversations often drifted off-task or left students feeling overwhelmed.

# AI Supports Learning When Aligned to Student Needs and Instructional Goals

> **Did the AI help students learn?**

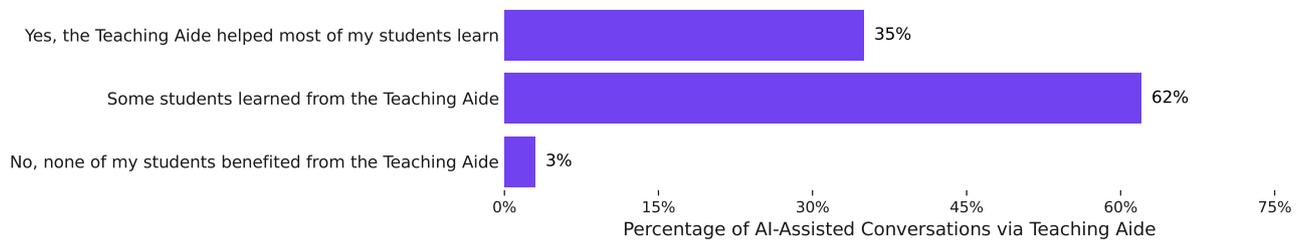

Note: Responses collected from 18 teachers. Weights normalized to 1 for teachers who reported multiple classes.

When asked whether AI-supported conversations enhanced student learning, teachers reported largely optimistic outcomes. 35% of teachers indicated that AI meaningfully supported student learning, serving as a helpful addition to traditional teacher–student interactions. Another 62% of sessions suggested that students were learning with AI, though teachers noted mixed results across the full class. Only 3% of sessions reported that students did not benefit from the AI's presence.

Teachers also described considerable variability in learning styles and outcomes, both within and across classes. For instance, a middle school math teacher observed that inquisitive, advanced learners frequently pushed the AI to explore topics far beyond the original assignment. Meanwhile, students who initially struggled with the content became engaged after being reassured they could start with just one part of a question. Similarly, a high school language teacher noted that multilingual learners, who had previously hesitated to participate in traditional whole-class conversations, found their voice through AI conversations. After practicing with the AI, these students gained confidence and began contributing more actively to whole-class conversations. These examples highlight the critical role of teacher scaffolding, design, and framing in transforming AI from a novel curiosity into a meaningful and personalized learning partner.

## Learning Patterns Varied by Grade Level and Subject

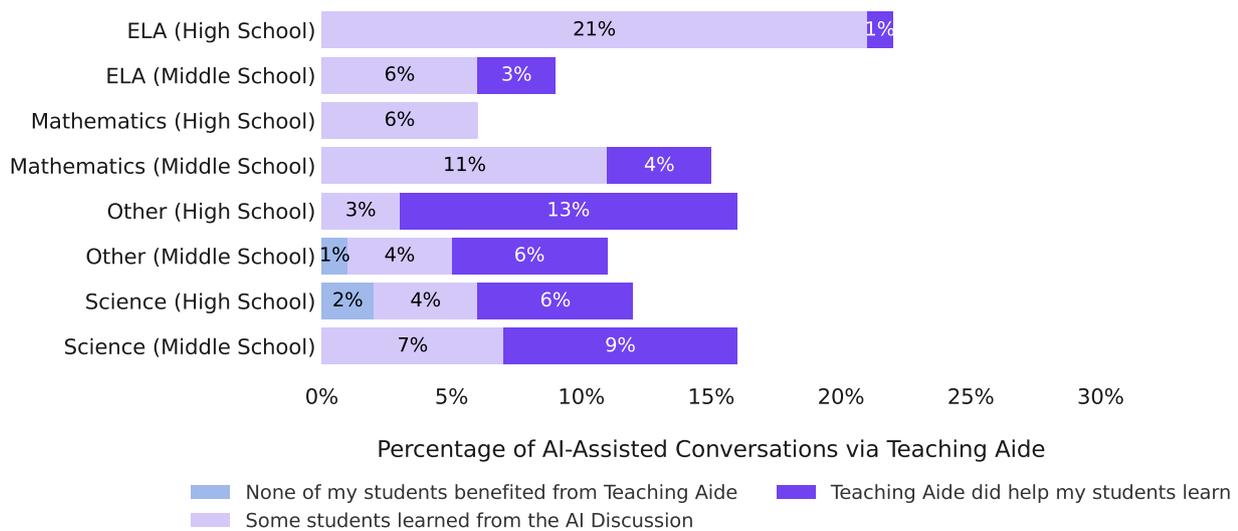

Percentage of AI-Assisted Conversations via Teaching Aide

- None of my students benefited from Teaching Aide
- Teaching Aide did help my students learn
- Some students learned from the AI Discussion

Learning outcomes reflected the similar grade-level and subject-area patterns observed in engagement. Science showed the clearest gains, with 6% of high school and 9% of middle school teachers reporting that Teaching Aide "did help my students learn," while high school "Other" subjects—such as Spanish, engineering, and interdisciplinary STEM—reported the highest levels at 13%. By contrast, ELA results were more modest: 21% of high school and 6% of middle school teachers noted that some students learned from AI conversations, but only 1% and 0%, respectively, reported strong gains. Younger students benefited most when teachers provided explicit scaffolds such as conversation starters and participation norms. High school students required additional support, including teacher modeling to build trust in AI and prompts that pushed conversations toward evidence-based reasoning. Subject-level differences also emerged: Spanish teachers leveraged AI for real-time, ACTFL-aligned proficiency feedback, shifting classroom focus from task completion to individualized skill development. High school English teachers used AI for prewriting and brainstorming, which improved essay quality. In science, AI-facilitated conversations supported inquiry and expanded opportunities for culturally relevant content. Mathematics and earth science classrooms produced more variable outcomes, with teachers pointing to challenges in generating accurate diagrams and in keeping AI aligned with the flexible, non-linear reasoning required for problem solving

# Teachers Were Confident in Using the Teaching Aide

**How confident do you feel using this feature again after their last implementation?**

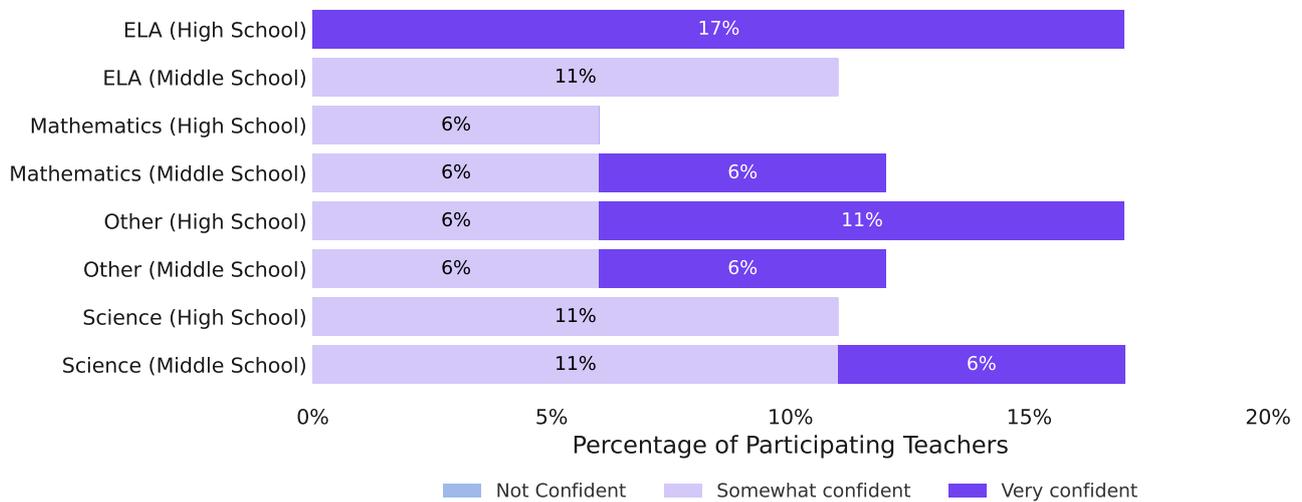

Note: Participant teachers who submitted at least one exit ticket for the Teaching Aide were included (N=18). For teachers who submitted multiple exit tickets, the confidence level from their final implementation was used. No teachers reported being 'Not Confident.'

In their final reported Teaching Aide session, over half of the teachers (56%) indicated they were "somewhat confident" in continuing to use the tool, while another 44% reported being "very confident." Although 2 teachers (11%) initially reported a lack of confidence after their first implementation, they became more confident in subsequent sessions. By the final implementation, no teachers indicated ongoing concerns about reusing the feature. Feedback emphasized the importance of adding monitoring tools and ensuring tighter alignment with instructional goals.

Overall, these findings suggest that student-AI conversations, while promising, require strong teacher mediation and developmentally appropriate scaffolds to succeed. Teachers who framed the activity as purposeful, integrated it into larger instructional goals, and maintained presence during the conversations saw the strongest student engagement and learning outcomes.

# Assessment and AI Grading as Formative Feedback Tools

During the pilot, teachers used the platform's AI-powered assessment and grading features. The Assessment feature enables teachers to generate or refine standards-aligned rubrics and automatically create feedback and grades for student work. When students submit their responses (text or handwritten), teachers can choose to enable instant access to rubric aligned AI feedback. The goal is to speed and scale formative feedback, elevate depth of student thinking, while preserving teacher authority over final grading judgments.[4]

Similar to the Teaching Aide activity, teachers submitted planning worksheets and exit tickets documenting their assessment implementations. Of the 19 teachers who participated in this phase of the study, 13 submitted feedback survey forms describing how they used the platform to design and administer assessments. This activity came closer to the end of the school year and several participating teachers cited time constraints for not providing written feedback. Nearly all participating teachers explored the platform's ability to generate rubrics, provide formative feedback, and support student revision. Teachers retained control over final grading decisions, using AI primarily as a formative aid. Across the pilot, educators found the tool particularly effective for streamlining feedback, increasing student revision opportunities, and reinforcing clear assessment criteria.[5]

---

[4] A demonstration of the Assessment and AI Grading tool is available on YouTube. This video incorporates learning from the codesign sessions that were used to improve the tool.
[5] For further details see: *Tian, Z., Liu, A., Esbenshade, L., Sarkar, S., Zhang, Z., He, K., & Sun, M. (2025, October). Implementation Considerations for Automated AI Grading of Student Work. In Proceedings of the Artificial Intelligence in Measurement and Education Conference (AIME-Con): Full Papers (pp. 9-20).*

## Subject and grade level of pilot implementation

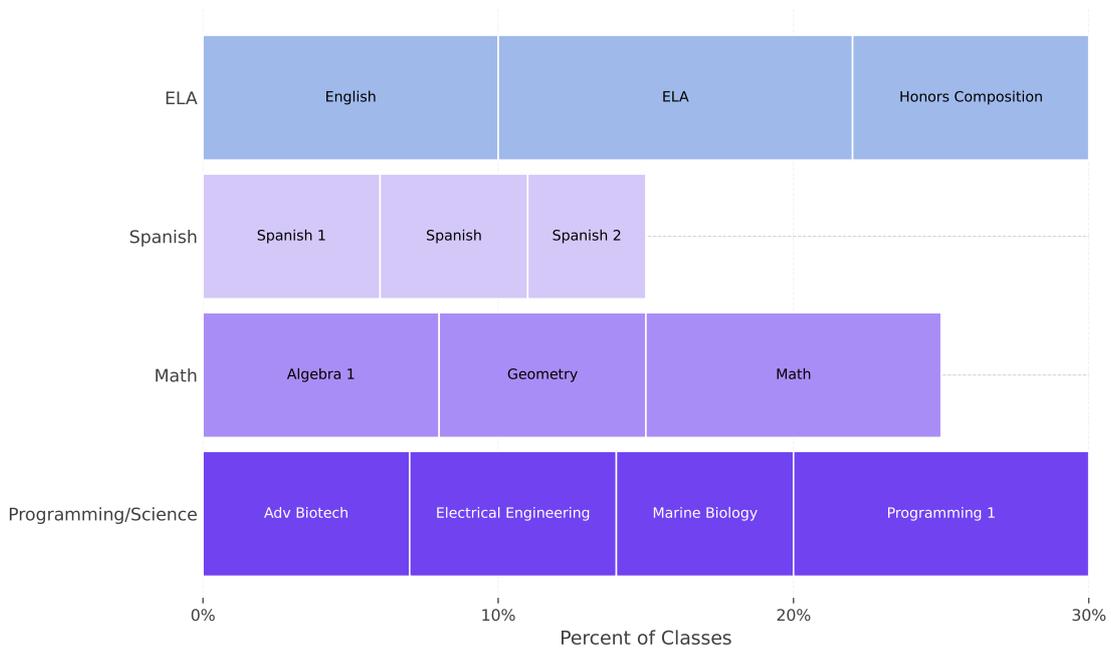

(a) Subject

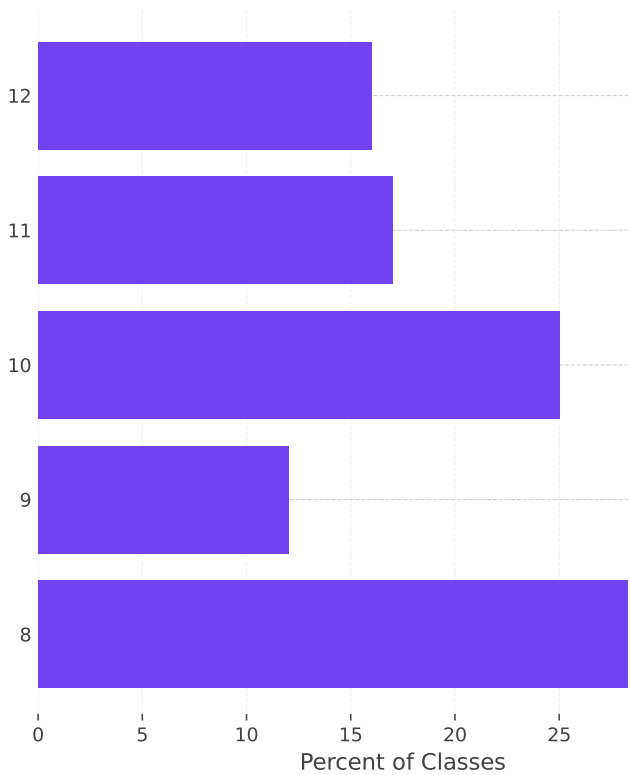

(b) Grade Level

Note: 13 teacher responses collected. Weights normalized to 1 for teachers who reported multiple classes.

The feature implementations spanned a range of subject areas: **programming and science courses (30%), mathematics classes (25%), Spanish language courses (15%),** and **English Language Arts (30%).** Classes were distributed across **grades 8–12.** Some teachers reported using the tool in multiple sections; because the individual teacher is the focal unit of analysis, responses were **weighted equally by teacher** (e.g., if one teacher reported a single math class and another reported two English classes, the study counts them as 50% math and 50% English).

## How teachers used Assessment feature

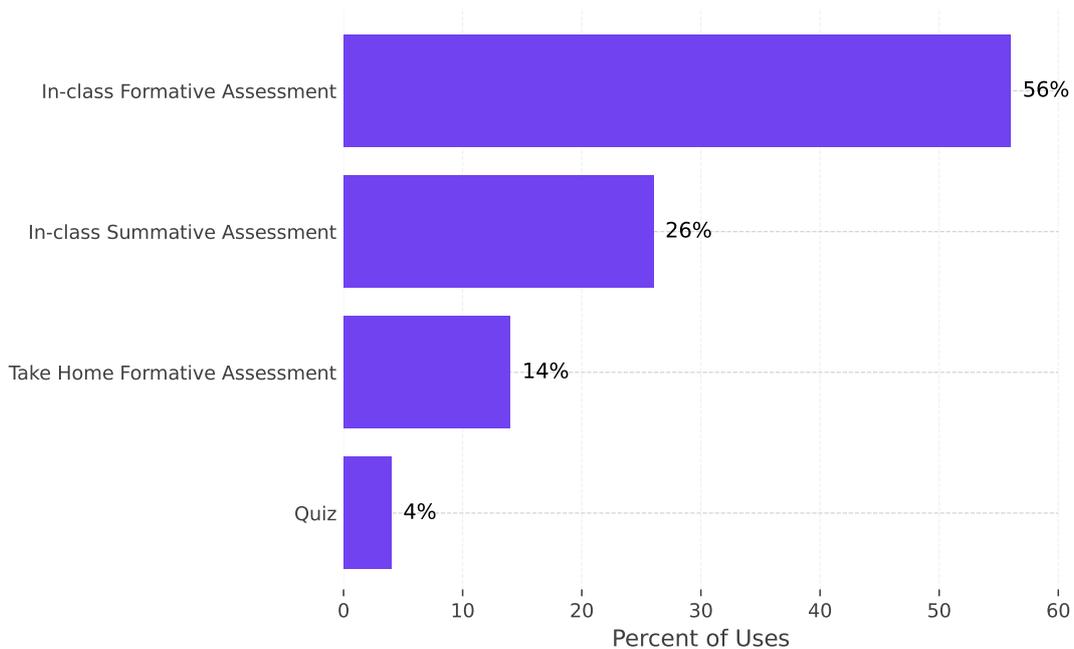

(a) Purpose

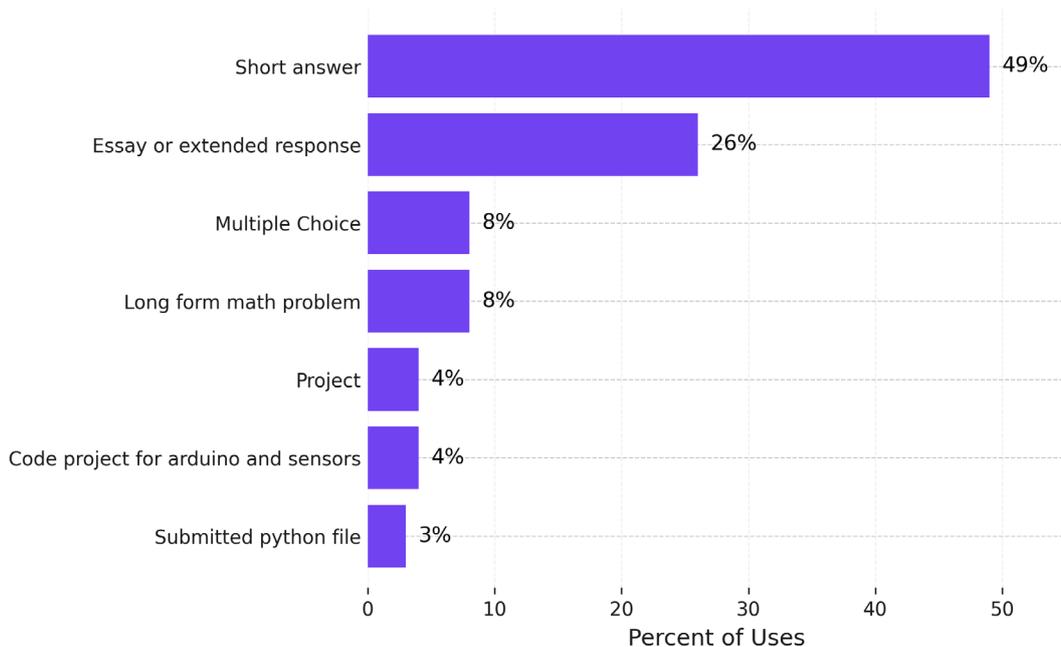

(b) Type

Although only 13 teachers completed the feedback survey, all 19 participating teachers used the platform to design and deliver at least one assessment. Over half (56%) reported using the platform for in-class formative assessments, while nearly half (49%) incorporated short-answer questions. These formats allowed teachers to gather real-time insights into student understanding and leverage AI-generated feedback to support learning during instruction.

## High AI Grading Coverage, with Student Engagement Shaped by Teachers

The platform dataset includes assessment activity in 58 distinct assessments created by the 19 participating teachers. Over the course of the pilot, these classrooms generated 936 student submissions across 33 classrooms, an average of approximately 1.8 assessments per classroom. Platform-generated usage log data were analyzed to understand how teachers implemented AI grading and feedback features across different subjects and school sites. The logs captured key details such as total student enrollment per classroom, number of submissions, AI-graded assessments, and the frequency of student resubmissions.

**Submission Patterns and Engagement.** Submission rates varied widely, with a mean submission rate of 54.8% (SD = 27.9%). While some classrooms achieved full participation, others showed near 0 submission rates, indicating variability in how assessment activities were adopted across contexts. This variation reflects both instructional choice and logistical constraints (e.g., class type, student access, timing).

**AI Grading Coverage and Automation.** AI systems graded the majority of submitted assessments. The median AI grading coverage was 92.2%, with many classrooms achieving near-total automation. In over 75% of classrooms, more than 80% of submitted student work received AI-generated scores. AI generated feedback was created when either a teacher or student (if allowed) clicked the 'AI grading' button. Both teachers and students can initiate AI grading to generate feedback and evaluation. This high rate of automated grading illustrates the system's capacity to streamline evaluation workflows at scale.

**Student Resubmission Behavior.** Students resubmitting the same assignment may indicate iterative learning or clarification efforts, was relatively infrequent but nontrivial. On average, 8.7% of students submitted work more than once, with a maximum observed rate of 66.7% in one classroom. This teacher described testing an intentional feedback cycle where students were asked to submit a 'first draft' to the AI and then incorporate its feedback into a final submission. While not ubiquitous, this behavior suggests some teachers and students leveraged the platform's capacity for revision and feedback loops.

# Table 2. Summary of AI Assessment Usage Across Classrooms

| Metric | Value |
| --- | --- |
| Unique Classrooms | 33 |
| Average Assessments per Classroom | 1.76 |
| Mean Submission Rate | 54.8% |
| Median AI Grading Coverage on Submitted Works | 92.2% |
| Average Resubmission Rate | 8.7% |

Note: Metrics are based on platform logs from 58 classroom-level assessment records across middle and high school implementations.

AI-powered grading was widely implemented across classrooms, with most student work receiving automated scores. Yet the variability in student engagement, along with uneven resubmission activity, reinforces a central finding from our qualitative analysis: teacher mediation remains essential to interpreting and contextualizing AI output. Teachers did not simply deploy automation, instead they integrated it into their classroom practices to balance speed with pedagogical intent.

# 60% of Teachers Found Grading Rubrics Useful and 57% Reported Helpful AI Feedback

Over 60% of the teachers indicated that they were able to use the AI generated rubrics in their classroom assignments. The majority indicated that they made minor changes to the rubric, indicating that they were not willing to fully accept the AI generated content without review and adjustment. Interestingly, no teachers indicated that they made major changes to the AI generated rubric. Only 7% of teachers indicated that the rubrics could not be used in their classroom - either because they needed major revisions or were simply not applicable. Roughly a quarter of teachers reported that they did not attempt to use the AI generated rubrics at all. Note that some teachers submitted multiple response forms for their different classrooms, the overall results are weighted so that each teacher has equal weight.

**Teacher survey responses to the quality of the AI generated rubric and whether it was necessary to make changes**

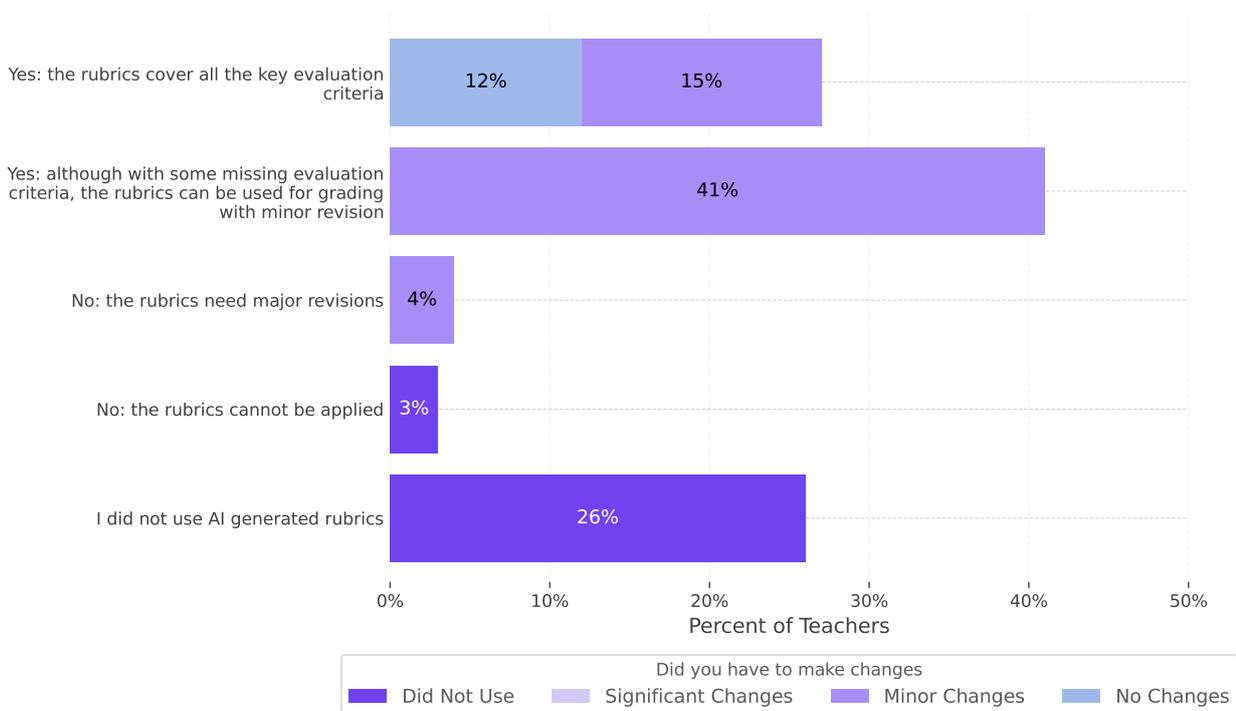

**Teacher survey responses to the quality of the AI generated assessment feedback**

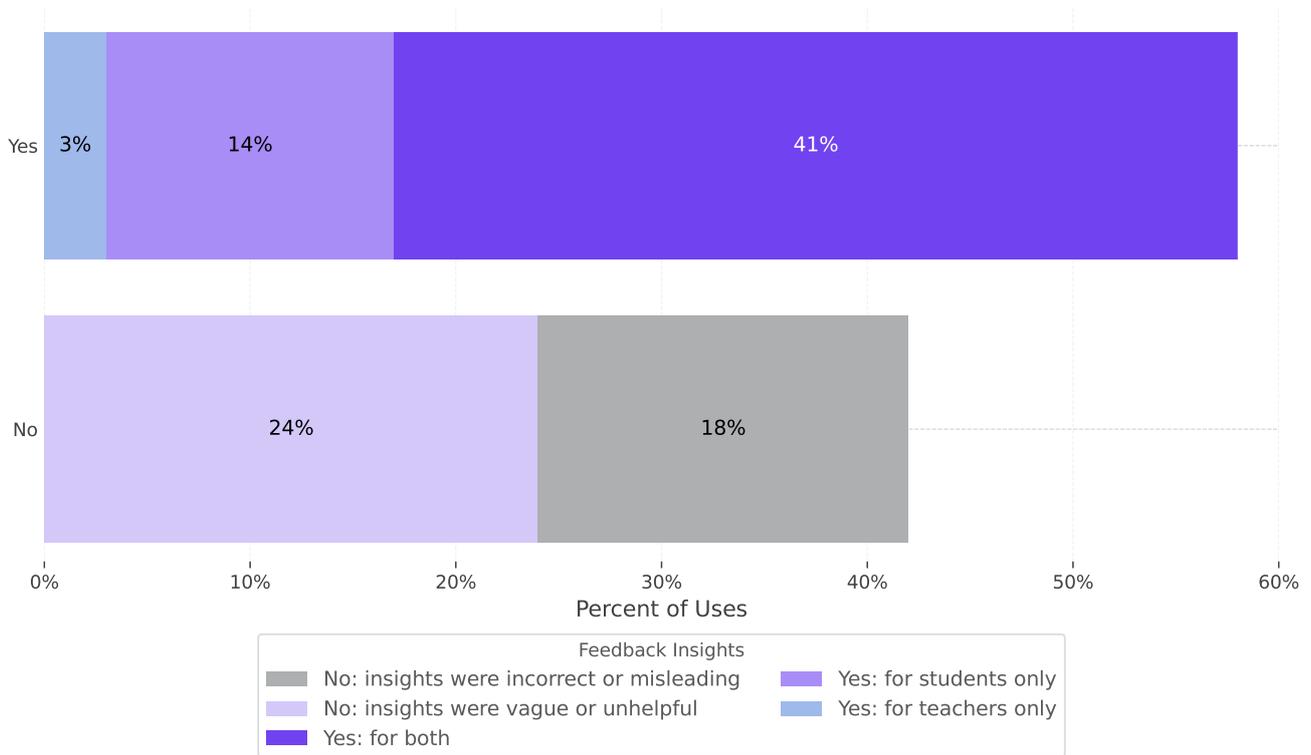

**AI Generated Feedback Quality.** 57% of teachers indicated that the AI feedback provided clear, actionable feedback for teachers or students, with 41% indicating that the feedback was useful for both teachers and students, 14% indicating that the feedback was only useful for students, and 3% indicating that it was only useful for teachers. 42% of teachers indicated that the feedback was not useful, with 24% indicating that the feedback was vague or unhelpful and 18% indicating that it was incorrect or misleading.

## AI Grading: Feedback as Formative Scaffold

Teachers consistently emphasized that the greatest value of the AI grading system lay in its narrative feedback rather than its numeric scoring. Although the platform provided an automated mechanism for assigning points, many teachers described the scores as inconsistent or misaligned with the expectations they had defined in their rubrics. In some cases, the AI applied different point scales across submissions or deducted points based on criteria unrelated to the assigned task. These inconsistencies led teachers to question the reliability of the numerical grades but did not undermine their appreciation for the written feedback.

By contrast, the AI's narrative responses were frequently described as useful, targeted, and well-aligned with formative learning goals. Teachers used this feedback to help students identify misconceptions, revise their work, and reflect on next steps. Many referred to the AI output as a helpful "first draft" of feedback—one that saved time and reduced the burden of writing initial comments. While the AI-generated scores were often disregarded or adjusted, the comments themselves were seen as actionable and pedagogically sound.

## Teacher Oversight Enables Trust and Personalization

Despite the efficiency of automated grading, students did not passively accept the feedback provided by the AI. Many questioned the fairness or accuracy of scores, especially when there appeared to be a mismatch between positive comments and low grades. This dynamic reinforced the importance of the teacher's role as mediator and interpreter. Teachers often stepped in to clarify or contextualize the feedback, helping students make sense of the AI's suggestions and re-establishing trust in the assessment process.

Teachers described this supervisory role not as a burden, but as an opportunity to personalize communication and reinforce key instructional messages. Some used the AI's comments as a base, rephrasing them in teacher voice or simplifying them for better student understanding. Others viewed the AI as a helpful assistant—a sort of digital teaching aide that provided immediate feedback, freeing them to focus on higher-order instructional decisions. In all cases, it was clear that teachers retained final authority over assessment and grading. Their presence ensured that students felt seen, heard, and fairly evaluated—something automation alone could not provide.

## Student Engagement is Mediated by Interface Design and Accessibility

Student engagement with AI-generated feedback varied, influenced not just by the content of the comments, but by interface design and accessibility. Teachers reported that some students, particularly those who were anxious or less confident, found value in receiving immediate feedback before sharing their work with peers. For these learners, the AI offered a low-pressure environment to revise and build confidence. However, some students were overwhelmed by the volume or complexity of the feedback and struggled to navigate the interface effectively.

Teachers noted that overly long comments could discourage engagement, especially for younger students or those with limited digital literacy. Technical issues, such as difficulties uploading files or navigating to specific parts of the feedback, further hindered the experience for some students. Others were put off by the platform's appearance, describing it as dated or unintuitive. These observations revealed that the success of AI-supported assessment depends not only on the quality of the feedback itself, but on the ease with which students can access, interpret, and act on it. A responsive, student-friendly design was seen as a necessary condition for broader engagement and equity in use.

# AI Feedback Supports Instructional Adjustments

Effective teaching depends on timely feedback that informs next instructional moves. In the pilot, Colleague AI's tools extended this process by capturing signals from student-AI interactions and distilling them into actionable insights for teachers. Teachers could see student curiosities, misconceptions, and points of struggle as they emerged. The AI Tutor, Conversation Summaries, and Student Growth Insights features work together to close the loop: students engage with the AI, the platform condenses and surfaces patterns, and teachers can use those insights to adjust instruction in real time. This section examines how these components contributed to more agile and responsive teaching, while highlighting the conditions under which they added the most value.

## AI Tutor as a Window into Student Curiosity and Struggles

In the pilot, teachers were encouraged to let students access the **AI Tutor** feature as needed. Unlike the Teaching Aide, which is structured around teacher-defined prompts and objectives, the AI Tutor provides an open-ended, freeform environment where students can independently explore topics of their choice, ask questions, and receive real-time AI responses. This flexibility allowed students to engage the AI in more authentic and personalized ways, often revealing their curiosities, misconceptions, and areas of academic struggle.

Although no formal evaluation was conducted specifically for the AI Tutor during the pilot, it shared the same underlying student-AI conversation architecture as the Teaching Aide. As a result, we collected indirect feedback from teachers whose students experimented with the tool. Many of the same challenges surfaced, particularly the length of AI responses, which some students found overwhelming or distracting.

One important difference between the AI Tutor and Teaching Aide is the level of structure. While Teaching Aide sessions are framed by the teacher with a specified topic, instructional objective, and expectations for both student and AI behavior, the AI Tutor operates without such framing. This makes the AI's ability to guide the conversation especially important. To address this need, the system prompt for the AI Tutor was revised to strengthen its Socratic teaching style and offer better progression through learning tasks.

As a result, a "suggested next step" feature[6] was added to reduce decision fatigue and support student thinking. After each AI response, two clickable suggestions now appear, offering students helpful directions they can use as the next turn in the conversation. This improvement aimed to scaffold open-ended inquiry while reducing distractions and helping students deepen or broaden their exploration in a more focused way.

---

[6] A demonstration of the AI Tutor tool is available on [YouTube.](#) This video incorporates learning from the codesign sessions that were used to improve the tool.

# Conversation Summaries Reduced Oversight Burden

For teachers, the challenge lies in making sense of unstructured student-AI exchanges at scale. To support this need, Colleague AI introduced two connected oversight tools: **Conversation Summaries** and **Student Growth Insights (SGI),** in addition to the full access to students' AI chat logs. Conversation Summaries provide a concise account of an individual student's interaction with the AI, highlighting the most important aspects of the dialogue without requiring a full transcript review. SGI aggregates patterns across many student conversations, presenting teachers with lists of the most common questions students asked and identifying recurring points of confusion.

Together, these tools form a feedback loop for instructional adjustment. Students interact freely with the AI Tutor; Conversation Summaries condense individual interactions for rapid review; and SGI highlights collective trends across the classroom. This cycle enables teachers to respond efficiently, whether by reteaching concepts, emphasizing specific skills, or designing follow-up activities that address identified gaps. Rather than manually scanning dozens of conversations, teachers can use these insights to keep instruction tightly aligned with demonstrated student needs while maintaining oversight of AI use in the classroom.

Pilot teachers reviewed over 140 student-AI conversation summaries and assessed them across five dimensions:

- **Accuracy** – factual alignment with the underlying conversation.
- **Relevance** – emphasis on the most important elements.
- **Completeness** – inclusion of all relevant information.
- **Clarity** – readability and ease of understanding.
- **Usefulness** – practical value for informing teaching.

Teachers scored each summary on a three-point scale: *No – several problems, Partially – some minor issues, or Yes – no changes needed.* Overall results were positive. Half of all conversation summaries received "Yes" ratings across all five dimensions, and only four summaries were rated "No" on more than one dimension. Clarity emerged as the strongest attribute, with 90% of summaries judged easy to understand. Ratings on accuracy, relevance, completeness, and usefulness were evenly distributed, with roughly two-thirds of summaries receiving "Yes" ratings on each dimension.

## Teacher Evaluation of Student-AI Conversation Summary

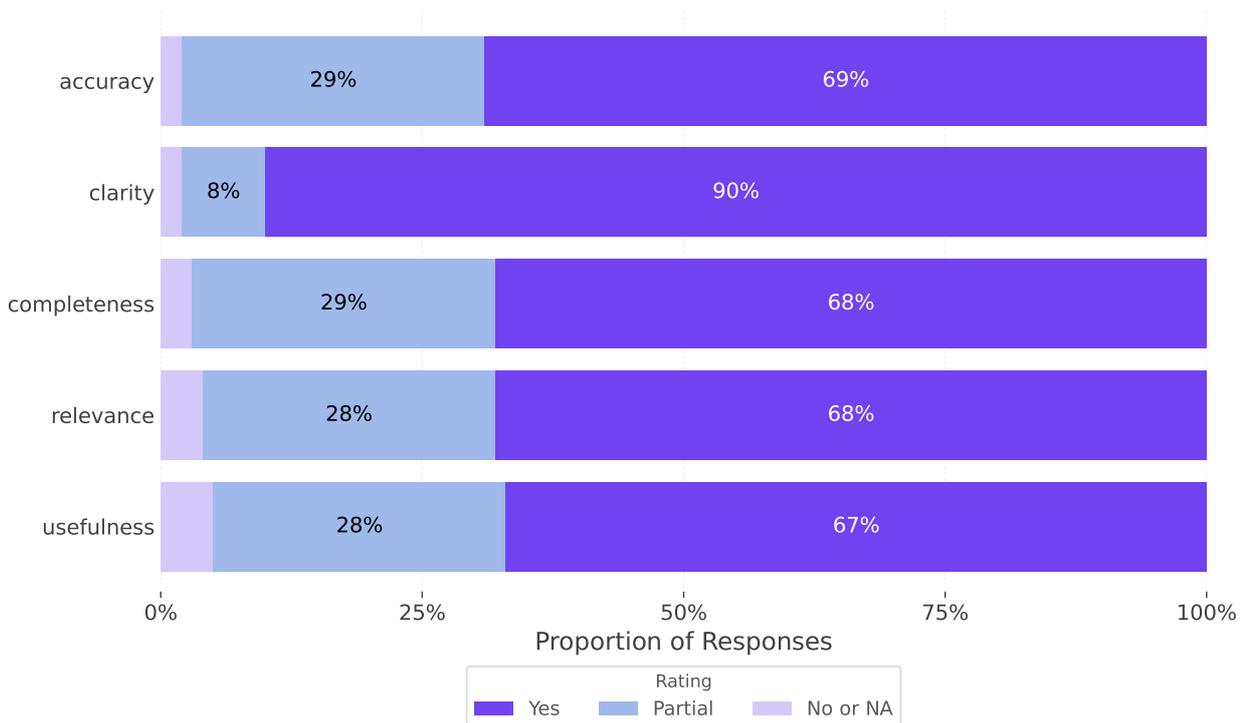

Teachers also identified recurring challenges. Summaries of very short or non-instructional student messages were sometimes less informative, and there were difficulties in representing cases where students expressed frustration toward the AI. Teachers also noted ambiguity in how the AI summary characterized student engagement within the interaction.

Automated conversation summaries are a promising approach for balancing oversight with efficiency in classroom use of AI. Teachers valued the summaries for providing clear, low-effort insight into student interactions, though refinements are needed to better capture edge cases and expressions of affect.

## Student Growth Insights Directed Real-Time Instructional Adjustments

The Student Growth Insights (SGI) feature provided teachers with aggregated summaries of student interactions on the platform, including lists of the most common questions students asked, recurring themes in student–AI conversations, and frequent mistakes identified in assessments. In interviews, teachers consistently reported that the common-question lists were the most valuable element of SGI.

By highlighting where students struggled or repeatedly sought clarification for the whole classroom and individual students, SGI gave teachers a clear, actionable direction of learning needs as they emerged in real time. Rather than conducting complex learning analytics, teachers could quickly identify patterns of confusion and adjust their instruction through reteaching, adding clarifying examples, or reinforcing specific skills.

This streamlined process with the Colleague AI Classroom across different features offered teachers a practical, low-effort mechanism for aligning instruction with demonstrated student needs, strengthening the feedback loop between student activity and instructional response.

# SGI Surfaces Actionable Patterns Strengthens, But Does Not Replace Teacher Judgment

During the pilot, 13 participating teachers submitted 139 total evaluations of the SGI feature, 15 at the class level and 124 at the individual student level. The analysis in this subsection uses the responses normalized by teacher to ensure equal weighting across educators regardless of how many submissions each teacher provided.

## Helpfulness of Student Growth Insights

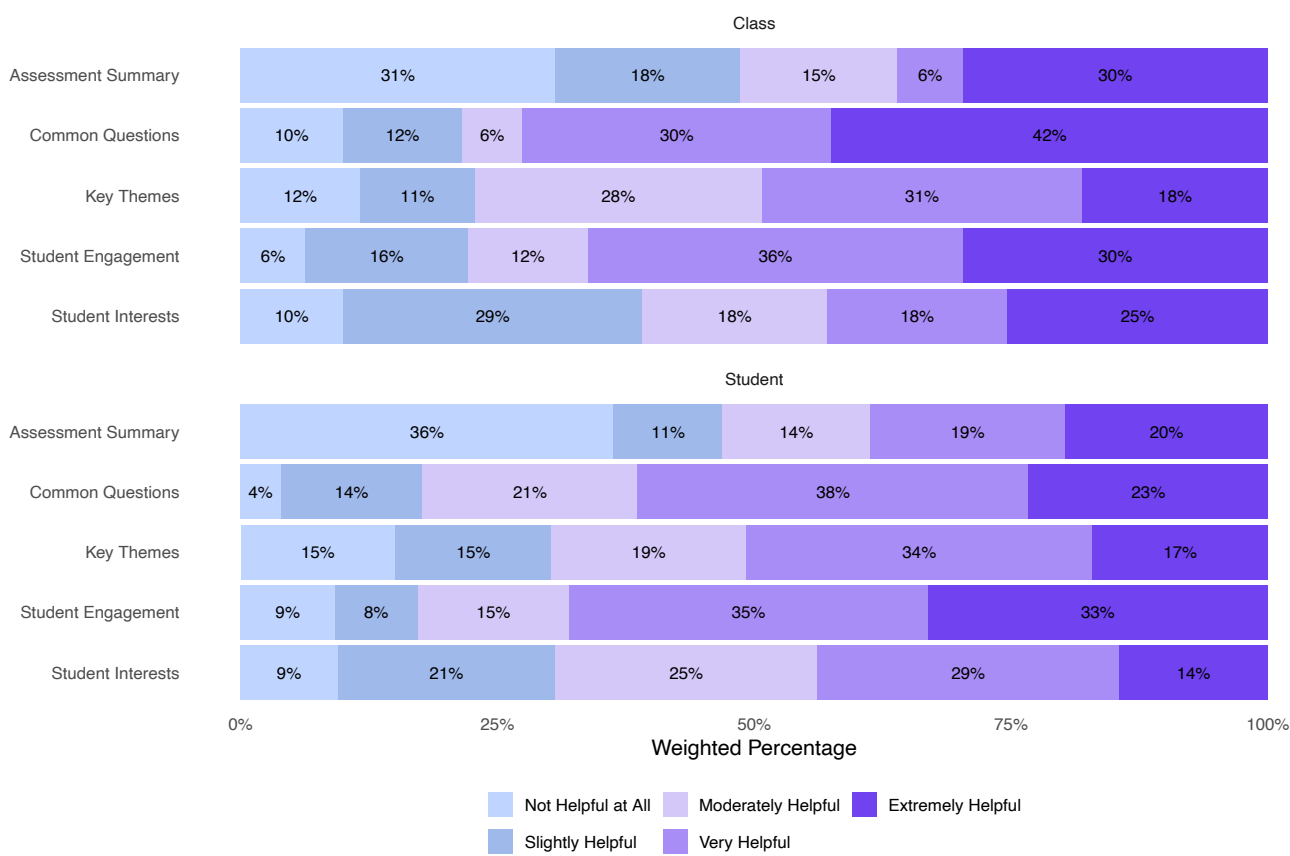

At the class level, SGI's most valued components were the "Common Questions" and "Key Themes" displays. 42% of responses rated the "Common Questions" section as extremely helpful, while 31% found "Key Themes" extremely helpful. Indicators of student engagement and interest were also frequently marked as moderately to very helpful, though they were less commonly cited as the most useful elements.

Student-level views followed a similar trend. "Common Questions" again received the highest marks for helpfulness, with 38% of responses rating them as very or extremely helpful. This was closely followed by student-specific engagement signals (35%) and key themes in conversation (34%), suggesting that teachers found these elements particularly useful for identifying patterns of confusion and guiding individualized support.

## Overall Evaluation of Student Growth Insights

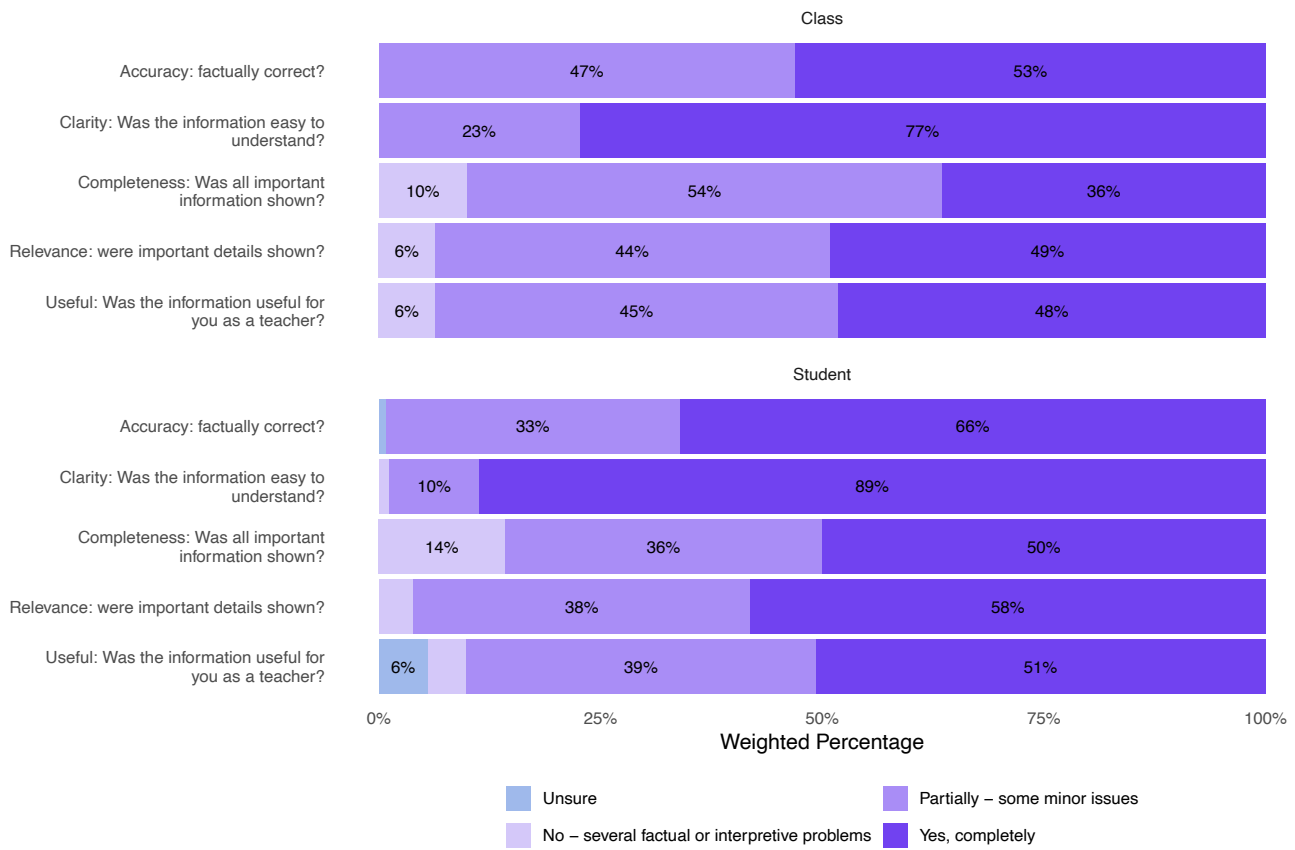

Teachers also rated SGI reports highly for their clarity, factual accuracy, and usefulness in practice. At the class level, all respondents (100%) indicated the insights were factually accurate, 53% said "completely accurate," and 47% said "partly accurate." 77% percent reported that the insights were easy to understand, and 90% believed the reports displayed all or most of the relevant information. In terms of classroom application, 94% found the class-level SGI reports useful to their instructional planning.

Student-level SGI reports received slightly more varied ratings but still demonstrated strong perceived value. 66% percent of responses identified the reports as completely accurate, and 89% found them clear and understandable. Just over half (51%) described the student-level reports as "very useful," with several teachers noting in feedback that interpreting individual SGI data sometimes required more time and attention than reviewing class-level trends.

Taken together, these responses suggest that SGI effectively surfaced actionable patterns, especially through the aggregation of common questions and themes, while maintaining a high standard of factual reliability and usability across both class and student levels.

> **The insights shown are based exclusively on the student's activity within the Colleague AI platform. How well do these insights reflect your overall understanding of your student(s)?**

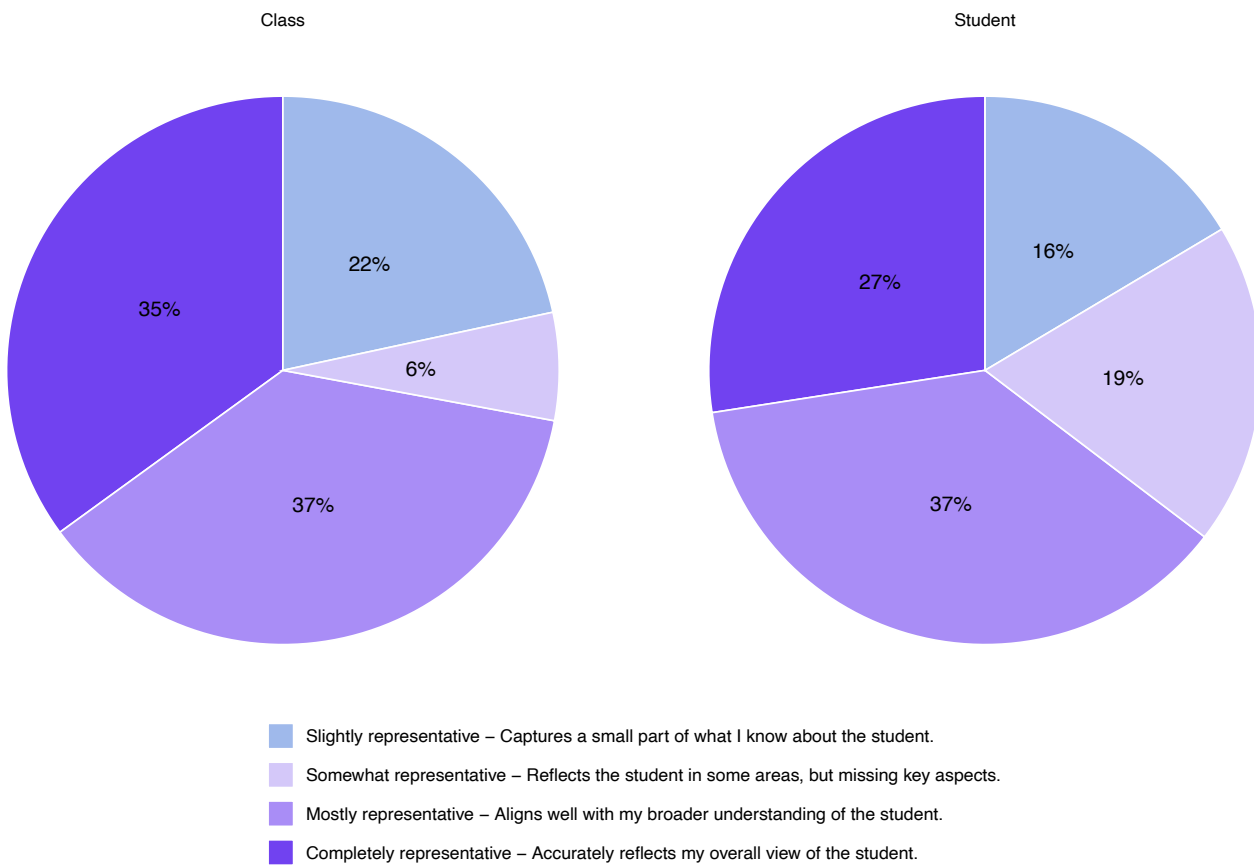

To assess the unbiasedness and relevance in perceptions provided by SGI, teachers were asked how well the platform's summaries reflected their broader understanding of student performance. This question was asked at both the class and individual student levels. As the growth insights only had access to information from the student activities in the online classroom - and did not access any external reports - we wanted to understand how well the platform based insights mapped to the teachers' broader perspectives of their classes and students.

At the class level, teachers expressed strong alignment. Nearly three-quarters (72%) reported that SGI reports were either mostly (37%) or completely representative (35%) of their classroom understanding. Only 6% felt the summaries were "somewhat representative," and 22% said they captured only a small part of what they knew, suggesting that, for most teachers, class-level trends matched their existing mental models of where students were excelling or struggling.

At the student level, responses were slightly more distributed, though still favorable overall. A combined 64% of responses rated SGI reports as mostly (37%) or completely representative (27%) of their individual understanding of a student. A higher share of teachers (19%) selected "somewhat representative," and 16% indicated the insights captured only a partial view. Several teachers shared that the largest discrepancy occurred for curious students that used their AI conversations primarily to test the AI and probe its capabilities, instead of engaging with it as a learning tool - the AI tended to report that these students were less engaged and in need of additional academic support.

These findings suggest that while SGI provides a solid baseline for interpreting student progress, teachers see it as complementary rather than exhaustive, most accurate when triangulated with other observations, but rarely contradictory to their own assessments. This positioning reinforces the feature's utility as a support tool for instruction, rather than a replacement for teacher judgment.

## Teachers See Potential in SGI-Customized Content Generation

The integration of SGI with content generation tools allowed teachers to immediately apply real-time learning data to the creation of lesson plans, rubrics, and instructional slides. Ten teachers evaluated this workflow, all of whom had used the SGI content-generation sidebar linking feature.

**Teacher Evaluation of Content Customization with SGI**

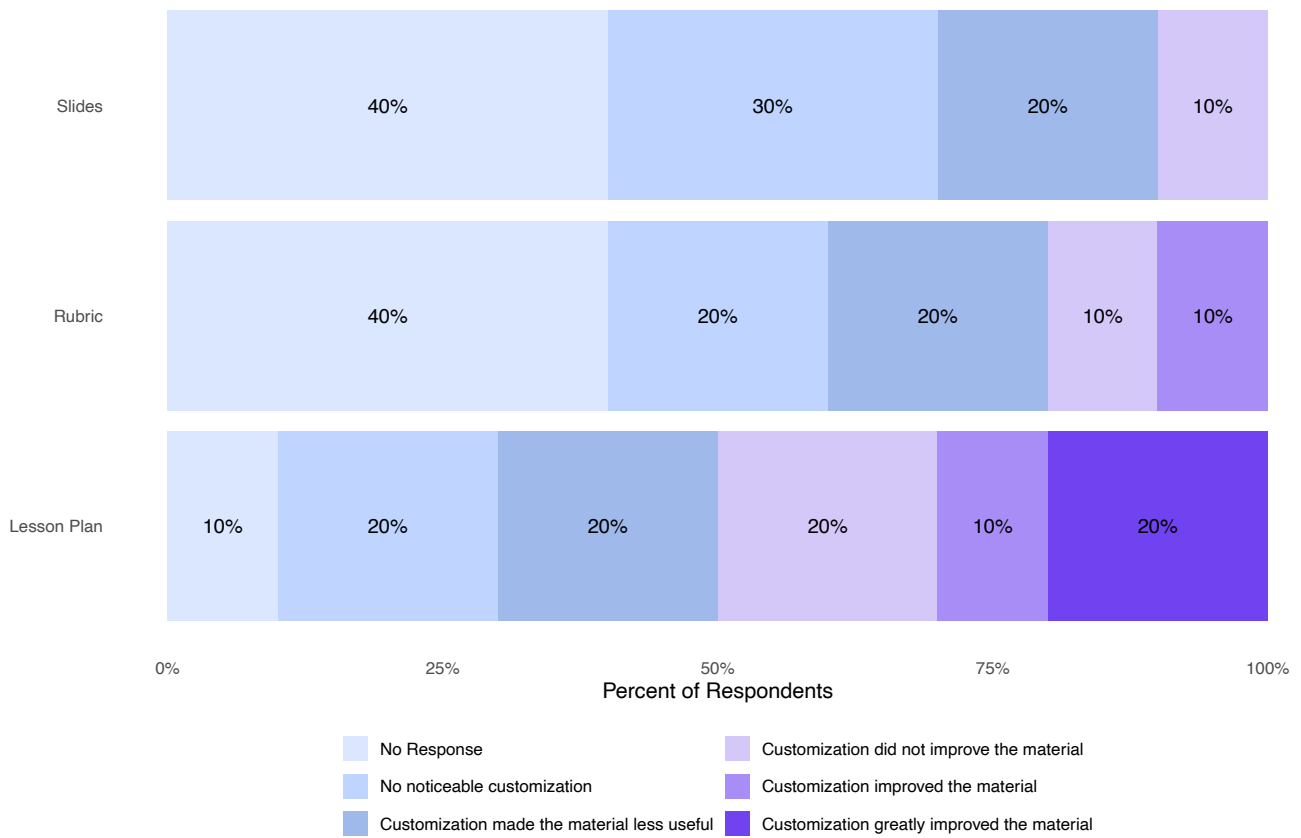

Among teachers who used SGI to generate and customize instructional materials, lesson plans received the most favorable feedback. Approximately 20% reported that customizing AI-generated lesson plans improved or greatly improved the material's usefulness. In contrast, fewer teachers found similar value in customizing rubrics or slides. Beyond lower overall usage for these two content types, rubric customization in particular showed a higher proportion of responses indicating "no noticeable customization" or that customization made the material less useful. This pattern suggests that rubric generation may require closer alignment with classroom expectations and more structured scaffolds to support meaningful teacher edits.

# Teacher Views on Continued Use and Integration of Content Generation with Growth Insights

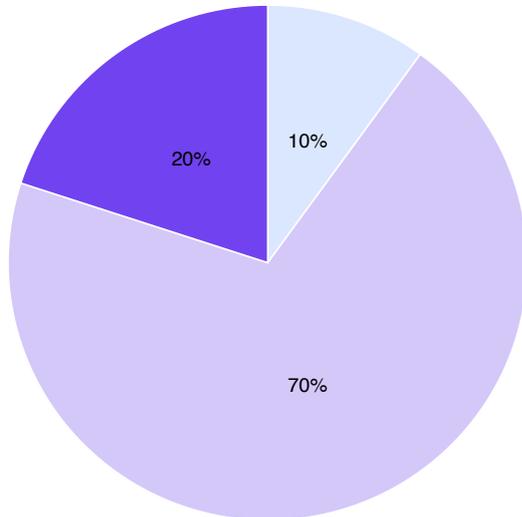

Will You Keep Using the Growth Insight Tool?

- No (10%)
- Sometimes (70%)
- Yes (20%)

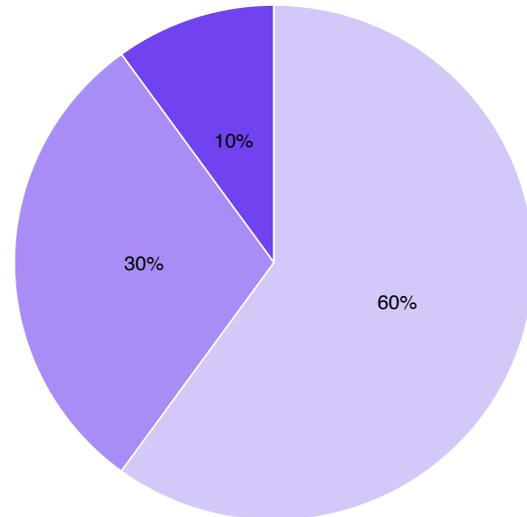

Do You Want All AI Chats to Use Growth Insights?

- Yes, in theory, but it needs to be developed more first (60%)
- Yes, but only if I can toggle it on or off (30%)
- Yes (10%)

When asked about future use, 90% of teachers indicated they would consider using the SGI content generation feature either sometimes (70%) or regularly (20%). However, support for broader integration across the platform remained conditional. A majority of teachers expressed the need for further development before full adoption, 60% preferred to see additional improvements first, and 30% wanted the ability to toggle the feature on or off depending on the context. Only one respondent (10%) supported full, always-on integration without controls. Notably, none of the teachers said that the AI Chats should not include student growth insights.

These results suggest that teacher interest in an SGI-informed content generation feature is strong. However, while some teachers found SGI-informed content generation useful, particularly for lesson planning, broad enthusiasm was tempered by concerns about customization quality, control, and fit for different instructional tasks. For this feature to be more widely adopted, teachers indicated that flexibility, refinement, and alignment to specific subject and classroom needs are essential.

# Insights from Teacher Co-Design: What's Needed for Effective AI Integration

## Teacher Reflections: Value, Friction, and Design Priorities

At the conclusion of the pilot, 14 participating teachers completed a post-study survey evaluating Colleague AI's effectiveness across four core instructional domains: developing instructional materials, grading and feedback, in-class instruction, and understanding student learning progress. Colleague AI was most highly rated for providing timely, meaningful feedback and surfacing insights into student understanding. Teachers noted that its feedback and assessment tools, as well as Student Growth Insights (SGI), were particularly helpful compared to other AI platforms.

> **Teacher-perceived effectiveness of Colleague AI in supporting specific teaching practices**

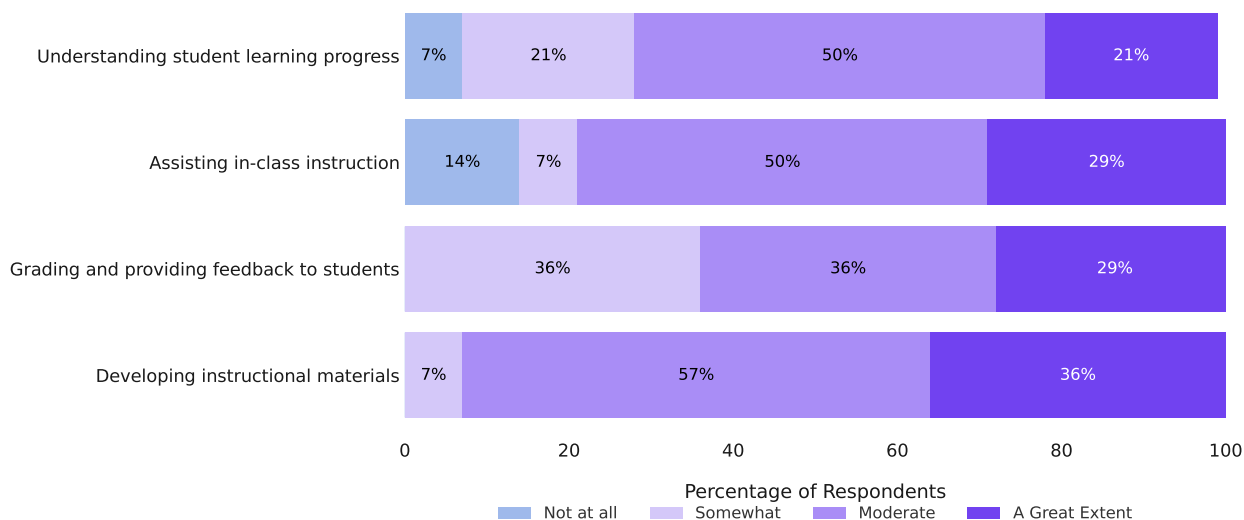

Question N = 14

Several teachers highlighted the platform's ability to facilitate structured student conversations and clarify learning needs. One **high school ELA teacher** in the post study survey reflected, "Colleague AI is superior in assisting me in class instruction, specifically in the [teaching aide] tools that we can modify and set up. Additionally, the insights function was incredible for supporting me in understanding student learning progress." Another added, "I'm not the teacher who just gives answers. I ask more questions, just like Claire AI from Colleague AI. That's why it works. We'd get along as teachers." Teachers valued this alignment with inquiry-driven instruction and the platform's scaffolding of academic conversations.

Colleague AI's fast turnaround on feedback was another frequently cited benefit. **An English teacher (Grade 11)** noted, "[Students] loved the prospect of getting a grade and feedback with such a quick turnaround, rather than waiting the 2-3 weeks that it usually takes me to grade their writing." **A programming teacher (Grades 9-12)** emphasized, "Without ClaireAI from Colleague AI, students would usually have to wait three or four days to hear back from me. With ClaireAI, they get instant feedback and can start making improvements right away—that's a big win."

In addition to speed, teachers appreciated that Colleague AI was not an unbounded chatbot, but rather a classroom-aligned tool. **A high school Spanish** teacher explained, "It's cool that Colleague AI already knows the ACTFL standards. We don't even have to input them. That makes it incredibly easy to align activities and feedback with national benchmarks without any extra work from the teacher." This sense of guardrails and instructional trust distinguished Colleague AI from general-purpose AI tools.

When asked how likely they were to recommend Colleague AI to colleagues (on a 1–10 scale), responses averaged 8 out of 10. Teachers who scored it highly appreciated the platform's structured discussions, formative feedback workflows, and insight tools. **A high school ELA teacher (score: 7)** remarked, "I would recommend it for the [Teaching Aide], AI Tutor, and Insights function, but the platform itself would be frustrating for some of my colleagues who are less comfortable with technology." **A science teacher (score: 9)** shared, "I have already recommended CAI to a few of my school colleagues, as well as my grad school classmates." **One ELA teacher (score: 6)** pointed to engagement issues: "The user interface is not intuitive... Students needed a lot of encouragement to engage with Colleague AI. It quickly became a chore for them." Others were more succinct: "Excellent resource." (Science, score: 10)

> **On a scale of 1 to 10, how likely are you to recommend Colleague AI to another educator?**

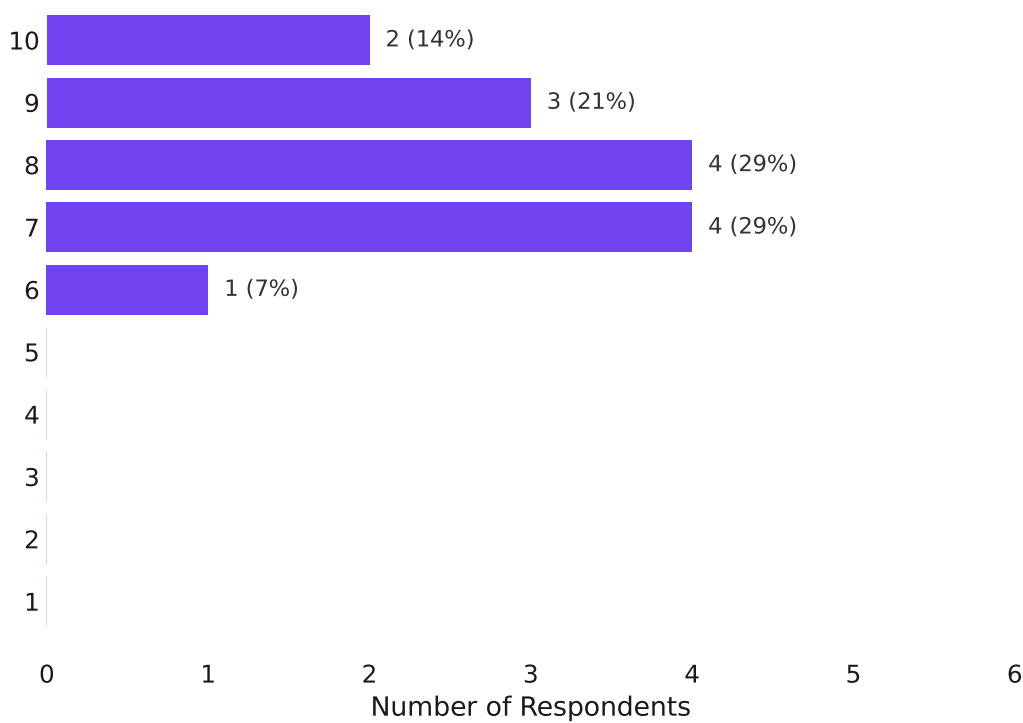

Question N = 14

# Subject Insights: Math

## Key Takeaways

- **Academic incentives boosted student engagement.** Math teachers tied AI conversations to participation grades, rubrics, or exit tickets, which increased effort and buy-in.
- **Scaffolding shaped success.** Teachers modeled math conversations, provided question stems, and set response-length expectations, while also prompting the AI to provide shorter, more focused replies.
- **Conceptual talk worked better than visuals.** Productive conversations centered on explanation, error analysis, and step-by-step reasoning, not diagrams or data tables where AI struggled.
- **Assessment feedback was useful, but scores were not.** Teachers valued AI-generated narrative comments to surface misconceptions and guide revision, but considered numeric scores unreliable or misaligned with expectations. They consistently reviewed and mediated AI output before using it with students.
- **Written reflections deepened understanding.** Having students restate math reasoning after AI exchanges reinforced learning and encouraged metacognition.

Math teachers primarily used Teaching Aide to support problem explanation, conceptual discussion, and formative assessment. Implementation frequency was notably high: one middle school special education math teacher facilitated 26 AI-powered conversations in a single week with the same group of students, using the tool to differentiate instruction and tailor support to individual learning needs. Overall, math conversations produced a bimodal distribution of engagement. Some students leaned into the tool, exploring new concepts deeply and actively engaging with the AI as a learning partner. Others contributed minimally or opted out altogether. A common concern from students was feeling overwhelmed by the volume of information and number of questions generated in AI responses. However, with simple reassurance—such as prompting students to "pick just one question to start," teachers found that many hesitant learners began to participate. Teachers consistently emphasized that their presence in the room, modeling how to engage with the AI, circulating to support, and reframing the task, was essential to ensuring that diverse learners could benefit meaningfully from the experience.

**Several barriers limited student engagement in math-specific AI conversations.** First, many students, particularly in younger grades, found the AI's explanations too long and its probing questions cognitively overwhelming. Teachers highlighted the need for more concise default responses and a toggle feature between "coach" and "explain" modes to help students manage information more comfortably. Second, the AI struggled with graphical reasoning tasks, such as interpreting diagrams, geometric figures, or tables, which limited its usefulness for geometry and data-based problems. Another common source of frustration was the AI's reluctance to provide direct final answers, especially in comparison to tools like ChatGPT. In these cases, how teachers framed the AI's role, emphasizing it as a tutor for thinking rather than a shortcut to solutions, was critical. Student engagement also dropped when AI conversations were ungraded or loosely connected to course objectives, as they were often seen as optional. In contrast, participation improved significantly when conversations were tied to small graded deliverables such as exit tickets or reflection prompts. Finally, teacher confidence played a substantial role. Educators who were unsure how to design effective prompts or who had limited knowledge of the AI's capabilities sometimes used overly broad or vague questions, which led to meandering, less productive conversations. Many requested structured prompt templates and curated libraries of effective examples to support more focused and impactful implementation.

**Math teachers used AI-assisted assessment to provide quick feedback on student reasoning tasks and short written responses.** While the system helped streamline initial review, teachers found the AI's scoring to be unreliable or misaligned with mathematical expectations. As a result, they used the feedback formatively and retained full control over grading.

In mathematics classrooms, the AI was primarily used to assess conceptual understanding, explanation of procedures, and short constructed responses. Teachers appreciated the time saved by the AI-generated feedback, especially when it surfaced errors or misconceptions that could guide revision. However, some found that the quality of the comments didn't always match their teaching style. As one middle school math teacher (Grades 7-8) put it, "The time-saving is great. But only if the comments represent how I would actually respond to student work. Otherwise I have to re-do it anyway." Teachers did not rely on the AI to assign grades; instead, they reviewed all output, made corrections as needed, and used the tool to extend student revision opportunities, not to replace professional judgment.

# Subject Insights: Science

## Key Takeaways

- **Verification was central to trust.** Teachers cross-checked AI outputs and modeled source checking, guiding students to validate claims and evidence, and encouraging them to think critically about the information provided rather than accepting it at face value.
- **Inquiry goals anchored productive use.** Science teachers saw strongest engagement when AI was tied to lab prep, Claims, Evidence and Reasoning writing, hypothesis generation, or debate preparation.
- **AI expanded representation in science.** Students encountered a broader range of scientists and perspectives, which helped expand their sense of who contributes to the field.
- **Middle schoolers needed concise scaffolds; high schoolers engaged with extended reasoning.** Middle school students responded best to short outputs and sentence starters that kept inquiry manageable, while high school students engaged productively with longer, Socratic-style explanations that supported deeper exploration.
- **Assessment feedback clarified reasoning, but scores were unreliable.** Teachers valued narrative feedback for helping students refine claims and evidence, but scoring inconsistencies required teacher mediation and final authority.
- **Structured prompts and clear products kept inquiry on track.** Focused entry questions and defined exit deliverables, such as lab notes or draft claims, prevented drift.

Science teachers integrated AI conversations into lab preparation, research projects, conceptual exploration, and scientific argumentation. Engagement was consistently high when conversations were tied to inquiry-based learning goals, such as developing hypotheses or constructing claims, evidence, and reasoning (CER). Teachers reported that AI worked especially well when embedded within larger instructional routines, for example, supporting students as they prepared lab reports, generated debate positions, or explored unfamiliar scientific topics. One middle school teacher used AI to help students research famous scientists and found that it not only provided comprehensive background information but also surfaced culturally relevant figures, which expanded students' understanding of who contributes to science and helped them see connections between their own identities and the broader scientific community.

**Despite its promise, Teaching Aide also presented subject-specific challenges in science classrooms.** Teachers valued the AI's broad knowledge base but remained cautious about its accuracy, routinely cross-checking outputs with trusted sources and encouraging students to verify information independently. **These practices were not only essential for building trust in the AI, but also helped cultivate students' critical thinking skills by positioning them as evaluators of information, not passive recipients.** In these moments, teachers' own AI competency and ethical modeling played a critical role, shaping how students engaged with the tool and influencing the accuracy, relevance, and reliability of the content brought into classroom discussions. Similar to mathematics, the AI often produced long, Socratic-style explanations that were better suited to high school learners; middle school students, in contrast, required shorter, more digestible outputs and sentence starters to scaffold their inquiry. Without clear prompts or a defined product, some conversations risked becoming too abstract or disconnected from the lesson.

To address these limitations, **successful science teachers paired AI use with clear, inquiry-aligned objectives and embedded source verification practices,** such as prompting the AI to cite its sources, provide links to referenced materials, or limit its responses to a defined set of trusted resources. Students were guided to enter AI conversations with focused questions, such as forming a hypothesis or interpreting a result, and exit with a preliminary claim, draft explanation, or plan for further investigation. Teachers integrated AI outputs into lab notebooks and CER frameworks, using feedback cycles to help students validate and revise their reasoning. Many also used AI to broaden scientific representation by exploring global perspectives, case studies, and underrepresented scientists. Across all grade levels, teachers emphasized the importance of teaching students how to verify AI-generated content, with the explicit goal of supporting them as thoughtful, discerning users of AI, using it as a springboard for critical thinking rather than a source of final answers.

**Science teachers used AI grading to support CER writing, lab reflections, and open-ended science explanations.** While narrative feedback was praised for helping students clarify claims and reasoning, AI-assigned scores were often inconsistent, prompting teachers to treat the tool as a revision support rather than a summative grader.

In science classrooms, the AI was frequently applied to analyze students' written arguments. Teachers found it particularly effective in helping students express scientific ideas more clearly and organize evidence logically. However, scoring issues were common. As one marine biology teacher (Grades 10-12) reported, "The tool scored some students out of 20 points and others out of 10, when I had specified the assessment was worth 10 points." Others noted that students were confused when feedback and grades didn't align. One high school science teacher (Grade 11) shared, "My students were confused why some feedback was so positive, yet the score was low. They came to me asking if the grade was accurate and what it really meant." Teachers emphasized the importance of mediation—reviewing, editing, and clarifying AI feedback to ensure alignment with content goals and student expectations. They also stressed that students must be taught to interpret AI feedback critically, understanding that the tool is a lever, not a replacement, for their own reasoning and judgment.

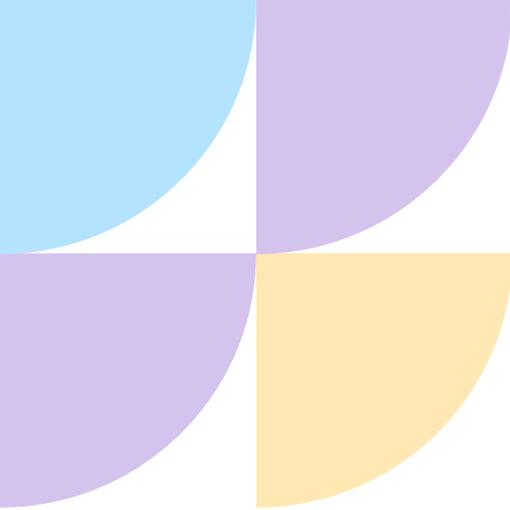

For questions or further information about this report,
please contact amplifylearn@uw.edu.

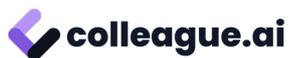
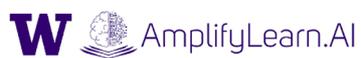
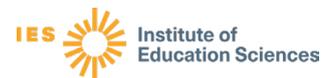

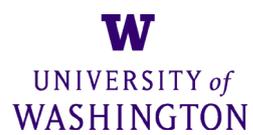
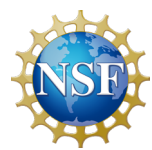

Acknowledgments: This work is supported by the Institute of Education Sciences of the U.S. Department of Education, through Grant R305C240012 and by several awards from the National Science Foundation (NSF #2043613, 2300291, 2405110) to the University of Washington, and a NSF SBIR/STTR award to Hensun Innovation LLC (#2423365). The opinions expressed are those of the authors and do not represent views of the funders.